\documentclass[twocolumn,astrosymb]{aastex701}

\newcommand{\healpix}{\texttt{HEALPix}}

\usepackage{amsmath}
\usepackage{url}
\begin{document}

\title{DELVEing into the LMC and SMC extended stellar periphery: tidal forces and the Stream's stellar counterpart}

\author[orcid=0000-0002-8093-7471,sname='Massana']{Pol~Massana}
% \altaffiliation{NSF's NOIRLab}
\affiliation{NSF NOIRLab, Casilla 603, La Serena, Chile}
\email[show]{pol.massana@noirlab.edu}  

\author[orcid=0000-0002-1793-3689, sname='Nidever']{David~L.~Nidever} 
% \altaffiliation{Montana State University}
\affiliation{Department of Physics, Montana State University, P.O. Box 173840, Bozeman, MT 59717-3840, USA}
\email{david.nidever@montana.edu}

\author[orcid=0000-0001-7827-7825, sname='van der Marel']{Roeland P.~van der Marel} 
\affiliation{Space Telescope Science Institute, 3700 San Martin Drive, Baltimore, MD 21218, USA}
\affiliation{Center for Astrophysical Sciences, The William H. Miller III Department of Physics \& Astronomy, Johns Hopkins University, Baltimore, MD 21218, USA}
\email{marel@stsci.edu}

\author[orcid=0000-0001-8251-933X, sname='Drlica-Wagner']{Alex~Drlica-Wagner} 
\affiliation{Fermi National Accelerator Laboratory, P.O.\ Box 500, Batavia, IL 60510, USA}
\affiliation{Kavli Institute for Cosmological Physics, University of Chicago, Chicago, IL 60637, USA}
\affiliation{Department of Astronomy and Astrophysics, University of Chicago, Chicago, IL 60637, USA}
\email{kadrlica@fnal.gov}

\author[orcid=0009-0000-9037-1697,sname='Oden']{Slater~Oden}
\affiliation{Department of Physics, Montana State University, P.O. Box 173840, Bozeman, MT 59717-3840}
\email{slater.oden@montana.edu}

\author[orcid=0000-0001-5486-2747,sname='de Boer']{Thomas~de~Boer}
\affiliation{Institute for Astronomy, University of Hawaii, 2680 Woodlawn Drive, Honolulu HI 96822, USA}
\email{tdeboer@hawaii.edu}

\author[orcid=0000-0003-1680-1884, sname='Choi']{Yumi~Choi} 
\affiliation{NSF NOIRLab, 950 N. Cherry Ave., Tucson, AZ 85719, USA}
\email{yumi.choi@noirlab.edu}

\author[orcid=0000-0001-5805-5766,sname='Riley']{Alexander~H.~Riley}
\affiliation{Lund Observatory, Division of Astrophysics, Department of Physics, Lund University, SE-221 00 Lund, Sweden}
\email{alexander.riley@fysik.lu.se}

\author[orcid=0000-0002-1594-1466,sname='Sakowska']{Joanna~D.~Sakowska}
\affiliation{Instituto de Astrof\'isica de Andaluc\'ia (CSIC), Glorieta de la Astronom\'ia,  E-18080 Granada, Spain}
\email{jsakowska@iaa.es}

\author[orcid=0000-0001-9438-5228,sname='Navabi']{Mahdieh~Navabi}
\affiliation{Department of Physics, University of Surrey, Guildford GU2 7XH, UK}
\email{m.navabi@surrey.ac.uk}

\author[orcid=0000-0003-1479-3059,sname='Stringfellow']{Guy~S.~Stringfellow}
\affiliation{University of Colorado Boulder, Boulder, CO 80309, USA}
\email{Guy.Stringfellow@colorado.edu}

\author[orcid=0000-0003-4341-6172, sname='Vivas']{A.~Kathy~Vivas}
\affiliation{Cerro Tololo Inter-American Observatory/NSF NOIRLab, Casilla 603, La Serena, Chile}
\email{kathy.vivas@noirlab.edu}

\author[orcid=0000-0002-8282-469X, sname='No\"el']{Noelia~E.~D.~No\"el}
\affiliation{Department of Physics, University of Surrey, Guildford GU2 7XH, UK}
\email{n.noel@surrey.ac.uk}

\author[orcid=0000-0002-7134-8296,sname='Olsen']{Knut~A.~G.~Olsen}
\affiliation{NSF NOIRLab, 950 N. Cherry Ave., Tucson, AZ 85719, USA}
\email{knut.olsen@noirlab.edu}

\author[orcid=0000-0002-6797-696X,sname='Cioni']{Maria-Rosa~L.~Cioni}
\affiliation{Leibnitz-Institut f\:ur Astrophysik Potsdam, An der Sternwarte 16, D-14482 Potsdam, Germany}
\email{mcioni@aip.de}

\author[orcid=0000-0003-4102-380X,sname='Sand']{David~J.~Sand}
\affiliation{Steward Observatory, University of Arizona, 933 North Cherry Avenue, Tucson, AZ 85721-0065, USA}
\email{dsand@arizona.edu}

\author[orcid=0000-0002-3204-1742,sname='Kallivayalil']{Nitya~Kallivayalil}
\affiliation{Department of Astronomy, University of Virginia, 530 McCormick Road, Charlottesville, VA 22904,
USA}
\email{njk3r@virginia.edu}

\author[orcid=0000-0001-6728-806X,sname='Gallart']{Carme~Gallart}
\affiliation{Instituto de Astrof\'{i}sica de Canarias, La Laguna, Tenerife, Spain}
\affiliation{Departamento de Astrof\'{i}sica, Universidad de La Laguna, Tenerife, Spain}
\email{carme.gallart@iac.es}

% Builders
\author[orcid=0000-0002-1763-4128,sname='Crnojevic']{Denija~Crnojevi\'c}
\affiliation{Department of Physics \& Astronomy, University of Tampa, 401 West Kennedy Boulevard, Tampa, FL 33606, USA}
\email{dcrnojevic@ut.edu}

\author[orcid=0000-0001-6957-1627,sname='Ferguson']{Peter~S.~Ferguson}
\affiliation{DiRAC Institute, Department of Astronomy, University of Washington, 3910 15th Ave NE, Seattle, WA, 98195, USA}
\email{pferguso@uw.edu}

\author[orcid=0000-0001-5160-4486,sname='James']{David~J.~James}
\affiliation{ASTRAVEO LLC, PO Box 1668, MA 01931, USA}
\affiliation{Applied Materials Inc., 35 Dory Road, Gloucester, MA 01930, USA}
\email{djames44@gmail.com}

\author[orcid=0000-0002-3690-105X,sname='Carballo-Bello']{Julio~A.~Carballo-Bello}
\affiliation{Instituto de Alta Investigaci\'on, Universidad de Tarapac\'a, Casilla 7D, Arica, Chile}
\email{jcarballo@academicos.uta.cl}

\author[orcid=0000-0002-9144-7726,sname='Martinez-Vazquez']{Clara~E.~Mart\'inez-V\'azquez}
\affiliation{NSF NOIRLab, 670 N. A'ohoku Place, Hilo, Hawai'i, 96720, USA}
\email{clara.martinez@noirlab.edu}

\author[orcid=0000-0003-0105-9576,sname='Medina']{Gustavo~E.~Medina}
\affiliation{David A. Dunlap Department of Astronomy \& Astrophysics, University of Toronto, 50 St George Street, Toronto ON M5S 3H4, Canada}
\affiliation{Department of Astronomy and Astrophysics, University of Toronto, 50 St. George Street, Toronto ON, M5S 3H4, Canada}
\email{gustavo.medina@astro.utoronto.ca}

\author[orcid=0000-0001-9649-4815,sname='Mutlu-Pakdil']{Burcin~Mutlu-Pakdil}
\affiliation{Department of Physics and Astronomy, Dartmouth College, Hanover, NH 03755, USA}
\email{Burcin.Mutlu-Pakdil@dartmouth.edu}

\author[orcid=0000-0002-6021-8760,sname='Pace']{Andrew~B.~Pace}
\affiliation{Department of Astronomy, University of Virginia, 530 McCormick Road, Charlottesville, VA 22904, USA}
\email{pvpace1@gmail.com}

\collaboration{all}{DELVE collaboration}

\begin{abstract}

The LMC and SMC’s influence on each other has produced a variety of stellar substructure in their periphery, which can help elucidate their interaction history.
We use the deep, multi-band DELVE-MC survey data to explore the stellar outskirts of the combined LMC+SMC system (XMC) and obtain new observational evidence of such interactions.
Using the old main-sequence turnoff region of the color-magnitude diagrams (CMDs), combined with custom star-galaxy separation and MW foreground models, we construct a map of the XMC stellar structures, extending 25\degr~from the LMC and 15\degr~from the SMC.  We recover all known structures, such as the northern LMC arm and SMC Northern Overdensity.  
We then compare the overall stellar density with a modeled symmetrical two-dimensional profile for each galaxy to enhance the different substructures. The model shows a clear asymmetry in the LMC profile, with a sharp drop-off in density along several directions near 8.6\degr~from its center. We interpret these density breaks as signatures of past SMC crossings. Additionally, 
we find a conspicuous enhancement of stellar density emanating from the SMC, in a broad range of directions that are in agreement with the HI gas Stream that was created by the interaction of the XMCs. We also note that this large spread of directions for the stream's stellar counterpart is in further agreement with recent independent findings. Based on our stellar density results, we suggest further searches for the Stream's stellar counterpart should focus along the gaseous stream path and the $-60$\degr~directions.

\end{abstract}



\section{Introduction}
\label{sec:intro}

The Magellanic Clouds (XMC) have long been the object of extensive scrutiny by the astronomical community. Their relative proximity has made them prime targets for resolved stellar population studies of dwarf and intermediate-mass galaxies, including, famously, the use of Cepheid stars to establish the first distance to the SMC \citep{leavitt_periods_1912}. The LMC and the SMC are the most massive satellite galaxies of the Milky Way (MW) and offer an unprecedented, close-up view of galactic interaction and hierarchical assembly. Understanding the structure and evolutionary history of the XMC is crucial not only for reconstructing the dynamical history of the Local Group but also for constraining cosmological models of galaxy formation. Historically, the morphological classification and structural analysis of the XMC relied on wide-field photographic plates. Pioneering studies in the 1950s \citep{de_vaucouleurs_studies_1955, de_vaucouleurs_studies_1955-1, de_vaucouleurs_studies_1957, Elsasser1958} mapped the primary optical extents of the Clouds, revealing their asymmetric features, the prominent LMC bar, and early hints of an extended, irregular periphery. However, the true extent of their stellar halos and the intricate debris from their ongoing interaction remained largely hidden due to depth and field-of-view limitations.

Because of the large apparent size of the XMC on the sky, searches for signatures of their outer halos have traditionally been very challenging and mostly focused on singular, small fields around their outskirts \citep{gallart_surface_2004,noel_old_2007,Saha2010,nidever_discovery_2011, nidever_exploring_2019, massana_smashing_2020}. In recent years, our understanding of the extended periphery of the XMC has been revolutionized by the advent of deep, wide-field photometric and astrometric surveys. The \textit{Gaia} mission has provided exquisite proper motions and parallaxes, enabling 3D kinematic mapping of the XMC system and its tidal debris \citep{gaia_collaboration_gaia_2021}. Concurrently, near-infrared campaigns like the VISTA Magellanic Clouds (VMC) survey have mapped the geometry of the inner and outer systems \citep[][and references therein]{Cioni2025}. Furthermore, deep optical imaging from campaigns utilizing the Dark Energy Camera (DECam)—such as the Survey of the MAgellanic Stellar History (SMASH; \citealt{nidever_smash_2017}), an independent search presented by \cite{Mackey2018}, the Dark Energy Survey (DES; \citealt{dark_energy_survey_collaboration_dark_2016}), and the Magellanic SatelLites Survey (MagLiteS; \citealt{drlica-wagner_ultra-faint_2016})—has unveiled a vast, complex stellar periphery. These observations have exposed expansive tidal tails, stellar rings, and far-reaching streams of debris that provide a record of the violent gravitational encounters between the LMC, SMC, and the MW \citep{mackey_10_2016, mackey_substructures_2018, besla_low_2016, belokurov_clouds_2017, choi_smashing_2018, choi_smashing_2018-1, martinez-delgado_nature_2019, cullinane_magellanic_2020, Oden2026}.

Beyond the tidal debris of the Clouds themselves, these wide-field surveys have uncovered a rich population of ultra-faint dwarf galaxies and remote star clusters residing within the extended XMC periphery \citep{Koposov2015, DrlicaWagner2015, Bechtol2015, martin_hydra_2015, Pace2025a}. The discovery of these satellites is of profound cosmological significance. In the current $\Lambda$ Cold Dark Matter ($\Lambda$CDM) paradigm, dark matter halos assemble hierarchically, meaning massive dwarfs like the LMC should bring in their own retinue of lower-mass subhalos when infalling into a MW-mass host \citep{Donghia2008, Li2008, Sales2011, Wetzel2015, Deason2015, SantosSantos2021}. The confirmation of a XMC satellite population \citep{Jethwa2016, Sales2017, Kallivayalil2018, Patel2020, Erkal2020} serves as a critical test for $\Lambda$CDM on the smallest mass scales, as it provides crucial observational constraints on the subhalo mass function and reframes our understanding of the ``missing satellites" problem by validating hierarchical accretion \citep{Deason2014, Deason2022, Cooper2025}.

Despite these advancements, fully mapping the extreme outer limits of the XMC periphery requires deep and contiguous sky coverage. In this paper, we present new, comprehensive maps of the extended stellar periphery of the XMC, utilizing deep photometric data from the DECam Local Volume Exploration (DELVE; \citealt{drlica-wagner_decam_2021, drlica-wagner_decam_2022}) survey obtained with the 4-m V\'ictor M. Blanco Telescope. This survey aims to complement existing archival DECam data to continuosly cover the Southern sky. Specifically, we trace the spatial distribution of the XMC using stars from the oldest main sequence turn-off (oMSTO), which motivated the depth requirements for the survey. While previous structural studies have frequently relied on brighter tracers such as red giant branch (RGB) or red clump (RC) stars, the oMSTO population offers distinct advantages for probing the ultra-low surface brightness periphery. oMSTO stars are intrinsically far more abundant, providing the necessary statistical weight in sparse regions, and they occupy a region of the color-magnitude diagram (CMD) that suffers significantly less from MW foreground contamination compared to the RGB and RC loci. Furthermore, stars from all ages and metallicities are present in this region of the CMD, which makes them less sensitive to population effects. Utilizing these optimized tracers, we derive updated stellar density profiles for the outermost regions of the LMC and SMC. We then analyze the distinct morphological features revealed by these maps, discussing their origins and how they could be tracing the complex formation, interaction, and dynamical evolution of the XMC over the last few billion years.

\section{DELVE-MC data}

All the data for this study come from observations taken with the Dark Energy Camera (DECam; \citealt{flaugher_dark_2015}) on the V\'ictor Blanco Telescope at Cerro Tololo (Chile). The data products used are produced by the MC group as a part of the larger DELVE survey collaboration \citep{drlica-wagner_decam_2021}. DELVE as a whole aims to complement existing and publicly available DECam observations with new ones designed to fill coverage gaps on the observable Southern sky. The survey was originally divided into three components: WIDE, DEEP, and MC, with the narrow-band CaHK MAGIC survey \citep{Chiti2026} added later. Each of these has a different goal for areal coverage and depth. For the MC in particular, the main goal is to achieve a continuous coverage around the XMC with similar depth to that of the SMASH survey \citep{nidever_smash_2017} in the \textit{g}, \textit{r} and \textit{i} bands. 

The MC catalog, used in this work, uses the same pipeline as SMASH. It covers a region of the sky defined as the combined area that goes out to 25 degrees from the LMC center and 15 degrees from the SMC center, which totals $\sim 2,200$ deg$^2$ of the sky. With this separate catalog, the aim is to focus on point-spread-function (PSF) photometry that better matches the crowded areas of the XMC main bodies, with the long-term goal of calculating star formation histories.

The overview of the MC data produced by the \texttt{delvered}\footnote{\url{https://github.com/dnidever/delvered}} pipeline, based on \texttt{DAOPHOT} \citep{Stetson1987}, is provided in Nidever et al. (2026, simultaneous with this work), where the processing and products are explained in detail. We will provide here a summary of the most relevant steps. 

\subsection{Forced photometry data}

In order to get to the depth we needed for oMSTO densities in the XMCs, we aimed for the deepest data possible. We achieve this by using only the forced photometry data coming from the \texttt{delvered} pipeline. In essence, in this step the pipeline produces a coadded image using all the exposures available (that meet the quality cuts) from the $g$, $r$, $i$ and $z$ bands, adequately weighted, and then performs source detection on that. Then, for each individual exposure, the PSF derived for that exposure is fit to the image pixels at the fixed positions (after accounting for appropriate astrometric transformations) of the sources detected on the coadds, with only the sky level and total flux as free parameters; this process yields so-called forced photometry, which is then calibrated into the native DECam filter system. We then correct the resulting magnitudes for extinction using \cite{schlegel_maps_1998} with the \cite{schlafly_measuring_2011} corrections.

There are two potential caveats of using the forced photometry data on the coadded exposures, as opposed to the full dataset. On the one hand, this means that you have to deal with measurements for sources that might otherwise have been below a certain S/N threshold for a given filter. Consequently, extra care must be taken with very faint sources as they could (and usually do) have very large uncertainties. Typically, for our footprint this threshold was at $g \sim 24$ mags or below, which is well fainter than the oMSTO. On the other hand, bright stars can be saturated at fainter magnitudes in the coadded image which would prevent them from appearing in the catalog. Our forced photometry, however, does not saturate below $g\sim17$. Therefore, this is not a problem for our study, as these stars are many magnitudes above the typical upper edge of the oMSTO ($g\approx22$).

\subsection{Star-galaxy separation and quality cuts}
\label{sec:stargalaxy}

\begin{figure}
    \centering
    \includegraphics[width=\linewidth]{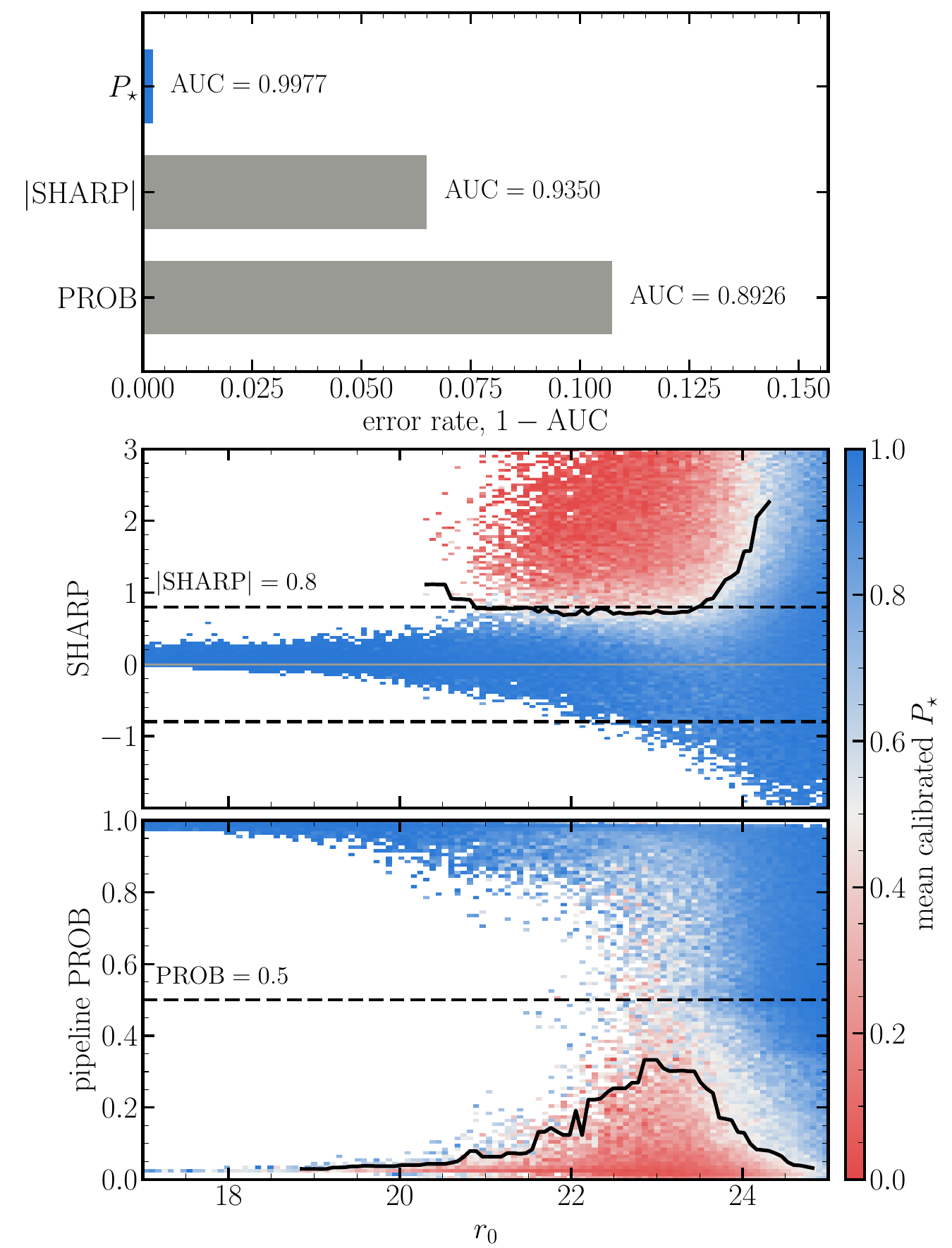}
    \caption{Summary of the \texttt{morphocloud} model capabilities compared to our benchmark HSC catalog labels. The top panel shows the AUC (Area Under the Curve) of different classification methods. A score of 1 means the classifier is perfect when compared to the benchmark labels. It shows how the stellar probability given by \texttt{morphocloud} outperforms the more traditional cuts that can be done with the \texttt{SHARP} and \texttt{PROB} values given by the \texttt{DAOPHOT} pipeline. The middle and bottom panels showcase this showing a two-dimensional histogram of a random sample of $\sim 3.89 \times 10^5$ sources outside the XMC galaxy cores showing the pipeline's SHARP and PROB parameters agains the $r_0$ magnitude, all color-coded by \texttt{morphocloud} stellar probability. The black solid lines showcase the median  P$_* =0.5$ boundary in the histogram.}
    \label{fig:model_summary}
\end{figure}

We applied additional cuts to the data to get the most accurate stellar density calculations. One quirk of the DELVE-MC catalog is that the reduction is done in ``bricks'' that overlap slightly. By setting \texttt{BRICKUNIQ}=1 we ensure that only unique sources are selected in overlapping areas.

After that, we found that using cuts based on \texttt{SHARP} and \texttt{PROB}, two standard \texttt{DAOPHOT} output features coming from the goodness of the PSF fit, are insufficient to remove galaxy contamination at the faint end, where the distinction between unresolved galaxies and stars becomes increasingly difficult. To that end, we used XGBoost \citep{chen_xgboost_2016} to develop a machine-learning model named \texttt{morphocloud} \footnote{\url{https://github.com/pmassana/morphocloud}}, to perform the star-galaxy separation. XGBoost, or eXtreme Gradient Boosting, is a package focused on delivering a high performance machine learning library that implements gradient boosting decision trees to make prediction models. We labeled stars and galaxies within our forced photometry catalog using a combination of existing catalogs, and in all cases, we used a cross-match radius of 0.5" with the DELVE-MC catalog. They are the following: 

\begin{itemize}
    \item \textit{Gaia} DR3 \citep{GaiaDR3_summary_2023}. It already has separate catalogs for stars and candidate galaxies as part of their data products, going down to G~$\sim21$. We did not use the QSO catalog, as those are essentially indistinguishable from point sources and often also overlap significantly with stars in color-color space. This is the most reliable labeling source we had, with the caveat that it only works for the bright end of our data.
    \item Hubble Source Catalog v3.1 (HSC; \citealt{Whitmore2016}) adds a source of labels at the faint end, though with very sparse coverage. We use the concentration index (CI), to separate stars ($0.8 < \mathrm{CI} < 1.2$) and galaxies ($\mathrm{CI}> 1.6$). Crucially, this catalog adds an important source of information inside the Clouds main bodies.
    \item Legacy Survey DR10 \citep{Dey2019} and DELVE DR3 \citep{drlica-wagner_decam_2021}. Although these are also produced using DECam images (often the same exposures), they have independent reductions and add a different source of information to star-galaxy separation. Legacy Survey uses Tractor to fit different types of surface brightness profiles to sources that are not point-sources. We label \texttt{PSF} type sources as stars and \texttt{EXP, DEV, SER} as galaxies. For DELVE DR3 we use the common Source Extractor cuts used in other works, ($|\mathrm{spread\_model}| + 3\times \mathrm{spreaderr\_model} < 0.003$) for stars and ($\mathrm{spread\_model} - 3 \times \mathrm{spreaderr\_model} > 0.005$) for galaxies. DELVE DR3 is the least reliable labeling source, so we also cross-match it to Legacy Survey to remove sources that have conflicting labels between them and finally we down-weight by 50\% all sources that only have a DELVE DR3 label.
\end{itemize}

The labeled dataset ends up containing around 169 million sources, split among 165 HEALPix pixels (nside=16). We use a 75/15/15\% training/validation/testing distribution of sources. Figure \ref{fig:model_summary} shows that the model achieves higher accuracy than \texttt{SHARP} and \texttt{PROB} cuts (top panel). For the comparison we use HSC as our ``source of truth'' as it is the catalog with the widest range of validity and remains completely independent from the DECam data. The middle and bottom panels show how \texttt{morphocloud} calculates stellar probability compared to the pipeline parameters across the magnitude range. We note that once $r_0>$23.5 mag the classification given by the machine learned model starts to lose accuracy and it is necessary to cut at very high P$_*$ to obtain a purer stellar sample. This is showing the intrinsic limitation of DECam data given its depth, at which point galaxies start to look very point-like and the classification is then challenging.

\begin{figure}
    \centering
    \includegraphics[width=0.95\linewidth]{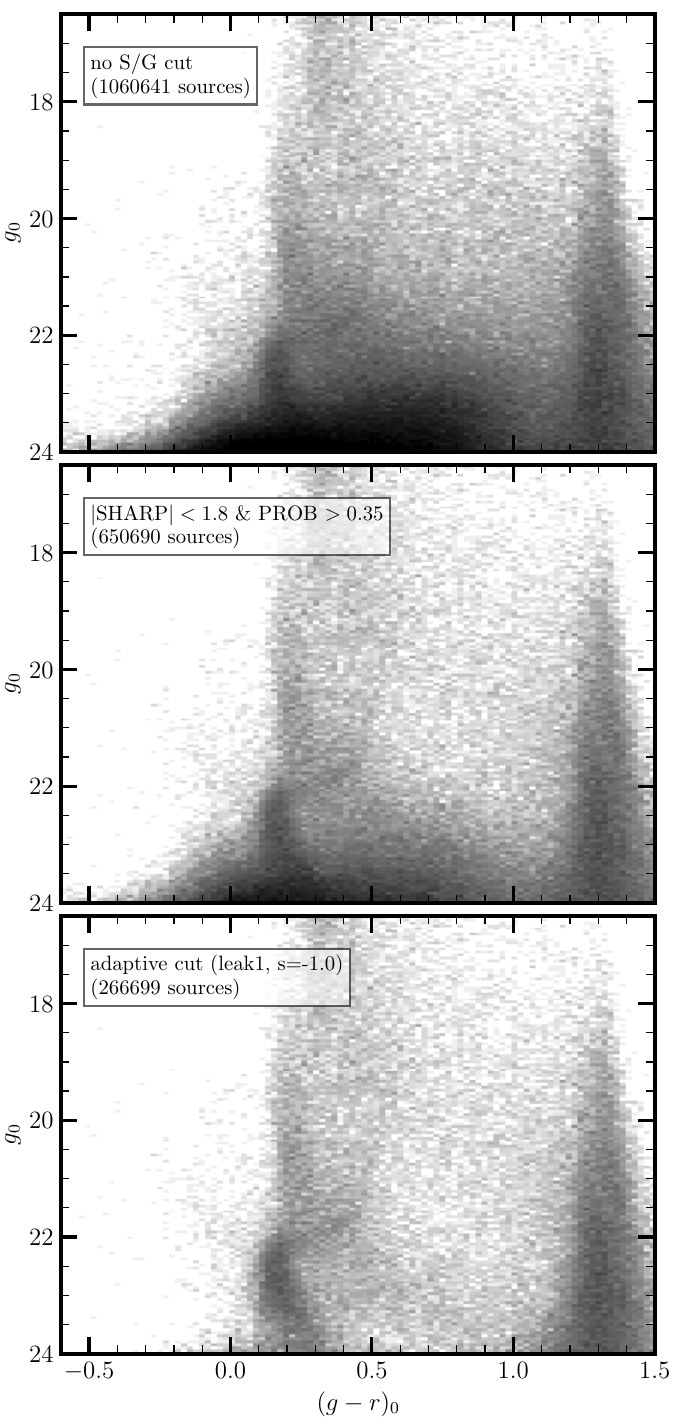}
    \caption{Color magnitude diagrams from a cone centered in (L, B)=($-10.4^{\circ}$, 4.1$^{\circ}$) in MS coordinates \citep{nidever_origin_2008} (or ($\alpha$, $\delta$) $\approx$ (55.95$^{\circ}$, $-65.38^{\circ}$)), and 1.5$^{\circ}$ in radius. All three panels use the same logarithmic color scale. \textit{(Top:)} CMD with no star-galaxy separation. \textit{(Center:)} Standard morphological cuts with the \texttt{SHARP} and \texttt{PROB} parameters that come from \texttt{DAOPHOT}. Significant contamination still remains at the faint end. \textit{(Bottom:)} star-galaxy cut using \texttt{morphocloud}, with a flat 0.8 stellar probability cut at $g_0>20$ mag and an adaptive cut that assures lower than 1\% galaxy leak, called ``leak1''. This becomes more noticeable at $g_0>23$ mag.}
    \label{fig:sg-separation}
\end{figure}

Once the model is trained, we also established an adaptive threshold of stellar probability that prioritizes stellar purity at the faint end, by ensuring a less than 1\% galaxy leak (also compared to HSC labels) that we name ``leak1''. The results can be seen in the bottom panel of Figure \ref{fig:sg-separation}, where we see the advantage of using the XGBoost model over using just the \texttt{SHARP} and \texttt{PROB} standard morphological parameters in the catalog.

\subsection{Spatial resolution}

One decision we needed to make is the spatial resolution to use when calculating stellar densities and making maps for this study. One starting point would be to base our calculations purely on the ``brick'' based system that the \texttt{delvered} reduction is based on, which are $0.25^{\circ}\times 0.25^{\circ}$ in size. This is a good resolution to show full maps of the whole footprint. However, to focus on specific stellar structures, we found that this resolution was too coarse to extract meaningful information. So instead we decided to resample our analysis to work in HEALPix with nside$=512$ ($\sim7$ arcmin resolution), which has a higher resolution but is still big enough to have good statistics on each element in the number of stars available. We are planning to made publicly available the HEALPix-level stellar density catalog as the main results of this work. 

\section{Milky Way model}
\label{sec: MW_model}

\begin{figure}
    \centering
    \includegraphics[width=0.95\linewidth]{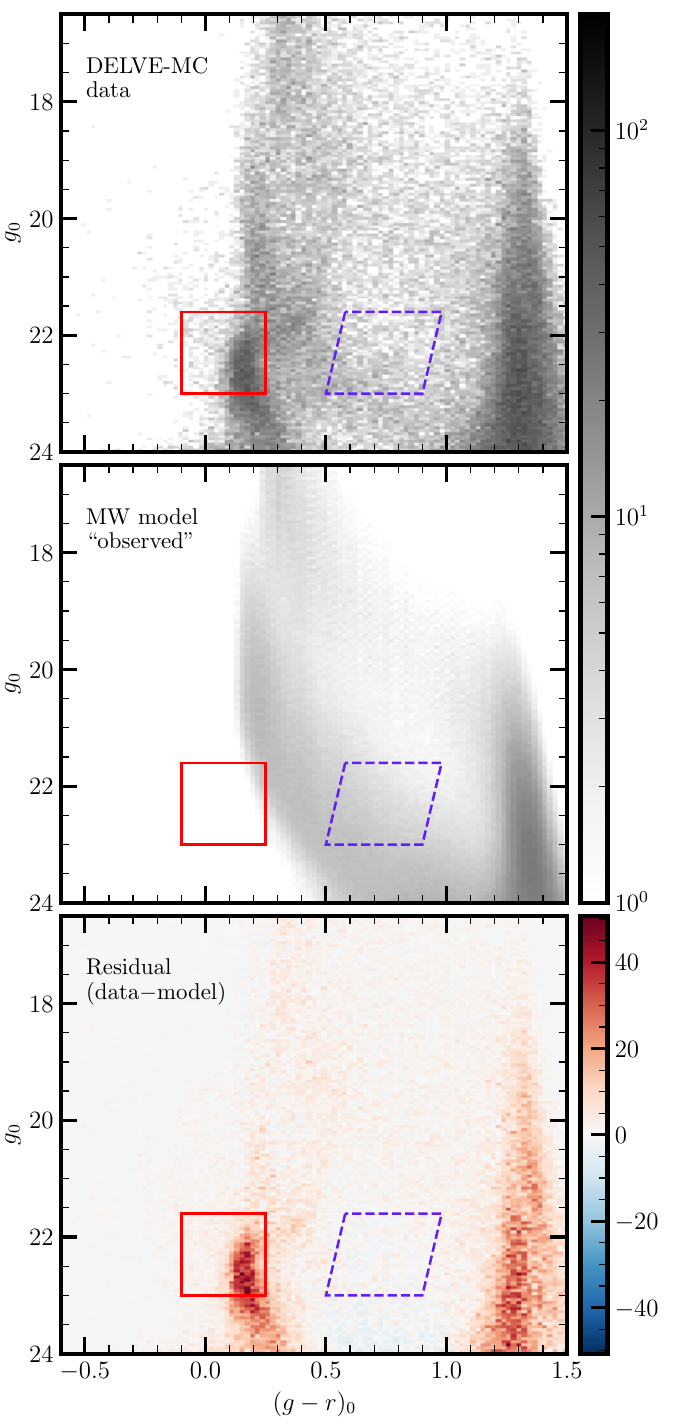}
    \caption{Color magnitude diagrams from a cone centered in (L,B)=($-10.4^{\circ}$,4.1$^{\circ}$) in Magellanic Stream coordinates \citep{nidever_origin_2008} (or ($\alpha$, $\delta$) $\approx$ (55.95$^{\circ}$, $-65.38^{\circ}$)), and 1.5$^{\circ}$ in radius. \textit{(Top:)} same panel as the bottom panel from Figure \ref{fig:sg-separation}, with forced photometry from DELVE-MC data, and star-galaxy separation as described in Section~\ref{sec:stargalaxy}. \textit{(Center:)} MW model with DELVE-like completeness and uncertainties to simulate observed values, as described in Section \ref{sec: MW_model}. \textit{(Bottom:)} Residual CMD after subtracting the model from the data. The dashed purple polygon indicates the area used to scale the model to the data. The solid red box in all the panels is the area used to calculate XMC member density.}
    \label{fig:data_vs_model}
\end{figure}

\subsection{Foreground contamination}
\label{subsec:foreground}

One of the key factors to be addressed when making claims about XMC stellar density is the membership classification of the stars in these two galaxies, and the ability to distinguish them from MW foreground sources. Being able to make this important distinction is one of the reasons why \textit{Gaia} data has been so successful in producing striking maps of the XMC stellar periphery. \textit{Gaia} astrometry provides a powerful means of distinguishing MW foreground stars from XMC populations because of their different kinematic signatures (see e.g. \citealt{belokurov_clouds_2019, gaia_collaboration_gaia_2021, massana_magellanic_2024, Oden2026}). Our photometry however is much deeper than what is provided by \textit{Gaia} and therefore we miss out on this crucial discriminator. Instead, having to rely exclusively on photometry makes the proposition of separating these populations much more difficult.

Our main instrument when trying to separate out different populations of stars has been using a CMD. Given prior knowledge of the location of different XMC stellar populations in a CMD, we can select those specific regions to reduce the impact of other MW populations. This approach however, is not perfect, as there are always MW stars distributed in those same regions. One area commonly used to pick out XMC stars is that occupied by their RGB stellar populations. But this overlaps the CMD area occupied by lower-mass main sequence nearby MW stars (closer to us than the XMCs). These are still relatively numerous even in crowded parts of the LMC disk. In order to reduce this MW contamination, it is better to use the CMD area around the oMSTO for XMC stars. The MW contamination in this area is mostly limited to stars that belong to the MW halo and are currently inside of the XMCs themselves as they pass through. This type of contamination, at the distance of the XMCs ($\sim50$~--~$60$ kpc; \citealt{pietrzynski_distance_2019, scowcroft_carnegie_2016}), should be minimal. The relative abundance depends on which part of the galaxy we are looking at, but for the main bodies of the galaxies, this has never been considered relevant enough to be taken into account. For the faint outskirts though, it could be important because in some areas of the footprint we get closer to the MW disk, which means we observe through the line-of-sight of denser parts of the halo.

\subsection{Implementation}

In this study, to account for MW foreground populations, we use a model created by T. de Boer (priv. comm.) that fits a set of MW components (thin and thick disk and halo) with parameters similar to Galaxia \citep{sharma_galaxia_2011} and then produces stellar populations using MESA isochrones and stellar tracks (MIST; \citealt{choi_mesa_2016}). For details about the model we point the reader to \cite{nidever_exploring_2019} and \cite{massana_smashing_2020}, both of which used the same MW model methodology. This time however, the model is expanded to cover the full DELVE-MC footprint and the components are fitted against DELVE-MC data. The results are a set of CMDs for each ``brick'' in the footprint that should represent the MW populations within it. These are purely theoretical with no observation completeness and uncertainties applied to them and are oversampled 100 times to increase statistics. So before we use them, we first randomly select a sample of stars of one hundredth the size of the original sample.

We derive a form of empirically extracted completeness and uncertainty relations for the DELVE data to apply to these models. Given that we do not yet have artificial star tests (ASTs) for DELVE-MC, we resorted to the SMASH ones that were the product of all the star formation history work done by that team \citep{ruiz-lara_large_2020, massana_synchronized_2022, sakowska_unveiling_2024}. These tests should represent fairly well the DELVE-MC data since the pipeline used in SMASH is also based on DAOPHOT and shares a significant part of the codebase. The tests were also run using the forced-photometry pipeline, and the data also come from the same instrument and filters.

Given the relatively unknown completeness characterization of DELVE-MC data, we use a simplistic function such as

\begin{equation}
 f(m) = \frac{1}{1+e^{\alpha(m-m_{50})}},
\end{equation}
where $m$ is a magnitude, $m_{50}$ is the point at which the completeness is at 50\% and $\alpha$ is the steepness of the distribution at that point. This function, similar to a Fermi-Dirac distribution, has been used before \citep{Harris2016} to describe the completeness of photometry data and should be well suited to our use case. During our analysis we noticed that the $\alpha$ parameter was most correlated with the density of the field, while $m_{50}$ depends on both density and the effective exposure times. So, based on the SMASH ASTs, we aimed to find a relation between density and $\alpha$, and $m_0$ and the peak of the luminosity function for the field, so that we could then later apply that to the DELVE-MC data. This yielded the following two relations:

\begin{multline}
        f(m_{LP}) = -3.60499453\times10^{-7} \rho^2 \\
    - 6.53120921\times10^{-4} \rho + 0.893346363
\end{multline}
\\
and
\\
\begin{equation}
    \alpha = 1.3991 + 1.0350 e^{-0.0113 \rho}.
\end{equation}
In both equations $\rho$ is the density expressed in stars per arcmin$^2$, while $m_{LP}$ is the magnitude at the peak of the observed luminosity function for stellar sources. These stars are selected based on simple \texttt{SHARP} and \texttt{PROB} cuts using the $g$ band. This gives a way to estimate the completeness at any given point of the footprint which can then be applied to the MW models. After that, we also need to apply photometric uncertainties. For that, again, we use SMASH ASTs to inform how these are typically behaved in a \texttt{PHOTRED}/\texttt{delvered} pipeline using DECam data. We find that the standard deviation of uncertainties at a given magnitude follows this approximate relation

\begin{equation}
    \sigma_m = 1\times 10^{-3} e^{0.7(m - m_{50} + 8)}.
\end{equation}
\\
Uncertainties are then computed from a normal distribution with zero mean and the corresponding $\sigma$. In the magnitude range of our study, the resulting uncertainties are relatively low.

Once both completeness and uncertainties were applied, the model still needs to reproduce any effect of the star-galaxy separation, or heterogeneous MW features like the Monoceros overdensity \citep{Conn2012}, which are not represented in the smooth model. 
In order to encapsulate all these different features that are not easily parametrized, we take a combined approach to solve them all empirically. Keeping in mind that we are mainly concerned here about reproducing the halo population as accurately as possible. To this end, we defined a small region in the CMD, which can be seen in Figure \ref{fig:data_vs_model} as a purple dashed polygon, that samples the observed MW halo population. We use the same magnitude range in this polygon as our XMC selection box to reproduce magnitude-sensitive effects like star-galaxy separation. And we also make this calculation for each HEALPix independently to encapsulate small scale variations. Star counts in this area are used to scale the whole model up/down to where the observations are, using a single scalar.

\section{Density maps}

\subsection{Calculation}

As mentioned in \ref{subsec:foreground}, our XMC stellar density calculation is based on the oMSTO region of the CMD. This is a method that was used before in SMASH by \cite{nidever_exploring_2019} and \cite{massana_smashing_2020}, where each study was focused on one of the Clouds. Here we are tackling the issue in a holistic manner and encompassing both Clouds. Hence, one of the first steps is to find a bounding box that can encompass the range of distances and different stellar populations that exist within the whole footprint. In Figure \ref{fig:periphery-map}, we can see in the side panels that the chosen bounding box is meant to encapsulate both the tighter LMC halo populations, and the SMC ones which have higher distance and metallicity dispersions. The lower limit is also informed by the observational completeness. We also explored dynamically adjusting the bounding box depending on the area, but noticed that this created artificial density boundaries across the sky area whenever we would switch from one to the other. Choosing a single bounding box provided a more homogeneous result. The chosen box is defined by $-0.10<(g-r)_0<0.25$ and $21.6<g_0<23$, and it encompasses a distance range of $\sim40$ to $\sim80$ kpc.

\subsection{Results}

\begin{figure*}
    \centering
    \includegraphics[width=\textwidth]{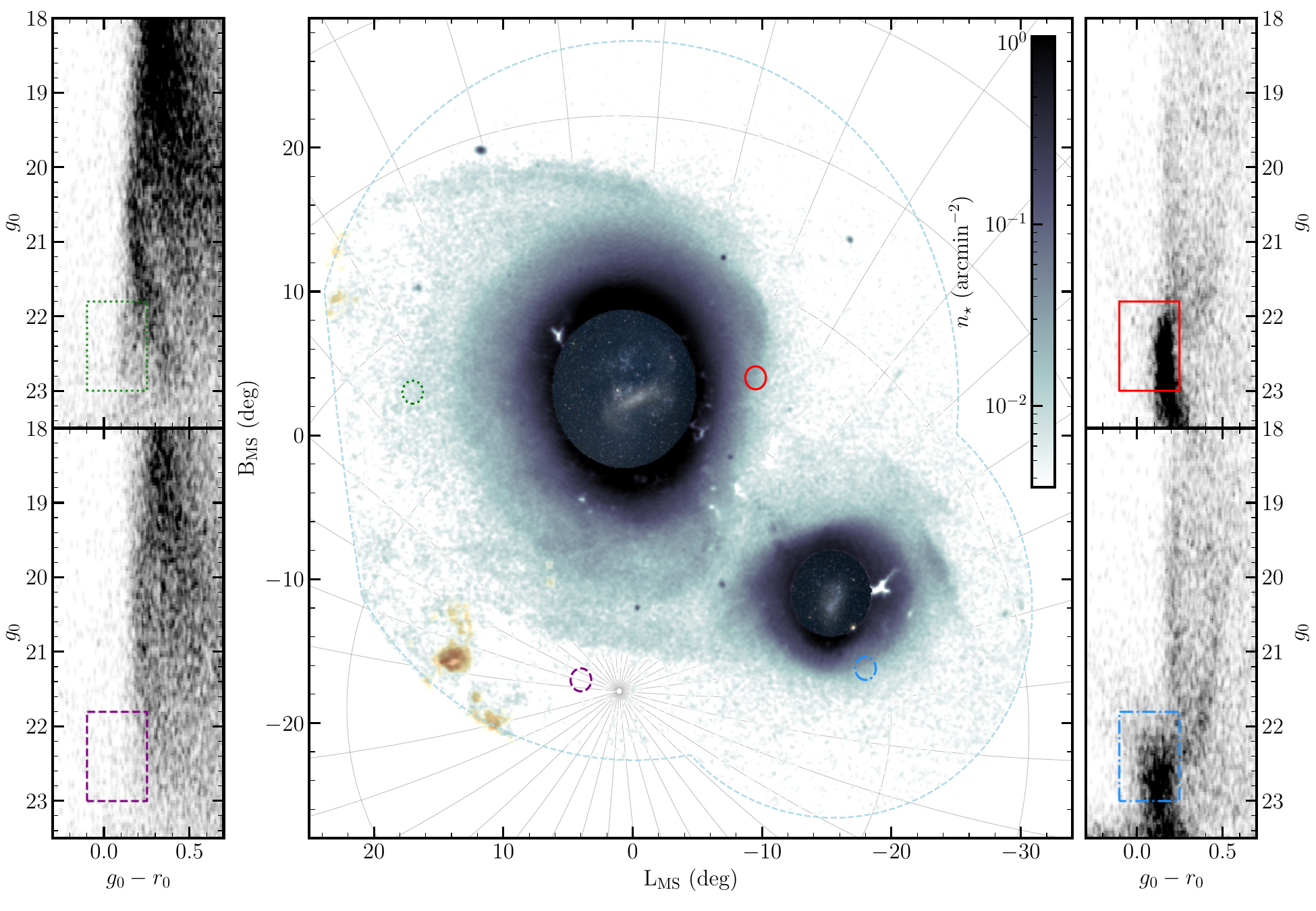}
    \caption{Map of the XMCs stellar periphery. \textit{Center:} Density of stellar sources around the XMC oldest MSTO in the CMD. The gridlines on the background show the Celestial coordinate system, while the light blue dashed line denotes the DELVE-MC footprint. Orange shaded areas are high extinction parts of the sky, where $E(B-V)>0.5$ mag according to \cite{schlegel_maps_1998}. The white gaps inside the Clouds represent small gaps in our coverage. The central parts of each galaxy are replaced by color images from the Mellinger full-sky survey \citep{mellinger_color_2009}. The different point-like and two stream-like overdensities are known satellites of the MW and the XMCs. \textit{Left and Right:} CMD of sources inside the small circular areas (0.8\degr~radius) in the central panel. The rectangles in the CMD indicate the area used to calculate the stellar density and the color indicated which part of the map they come from, matched by their color and line style.}
    \label{fig:periphery-map}
\end{figure*}

The main density map we are presenting in this work can be seen in Figure \ref{fig:periphery-map}. Thanks to the depth provided by the DELVE-MC data and the care we have taken in reducing MW contaminants, this gives us the clearest view of the far outskirts of this system yet. Already known features like the SMCNOD \citep[north/right of the SMC;][]{pieres_stellar_2017} and the LMC's Northern Arm \citep[north/top of the LMC;][]{mackey_10_2016} show up very clearly in this map (see also the annotated version in Figure \ref{fig:periphery-annotated}). The stellar halos of both galaxies are fully overlapping on the projected sky plane. While it is true that the line-of-sight distance of both galaxies is $\sim$10 kpc apart, the LMC disk is tilted towards the direction of the SMC \citep{van_der_marel_new_2002, choi_smashing_2018, Oden2026}, which itself has a very large distance spread \citep[$\approx \,20$ kpc;][]{nidever_tidally_2013, scowcroft_carnegie_2016}. It therefore seems likely that the periphery structures we are seeing have a physical overlap. This could help reconcile some results that appear to show either a connecting old bridge of stars \citep{belokurov_clouds_2017} or two overlapping stellar halos \citep{Jacyszyn-Dobrzeniecka2017}. This projected overlap appears much stronger in the direction of this speculated RR Lyrae old stellar bridge, with the LMC Hooks stellar substructure seeming to contribute to the connection of the two galaxies.

Along the Eastern side of the LMC it has traditionally been difficult to reliably confirm if debris that we are seeing comes from genuine XMC stellar populations or if it is MW contaminants. We have added four CMDs on Figure \ref{fig:periphery-map} to attempt to show that the vast majority of the debris is in fact from the XMCs, given our careful MW decontamination. The right hand side CMDs are there to show how the LMC and SMC have stellar populations that look distinctively different due to their different metallicities and line of sight distance spreads. The top-left CMD shows a clear LMC-like population inside the green dotted rectangle. The bottom-left panel has no distinctive XMC population inside the purple dashed rectangle, though there is seemingly a disperse low density population that does not show on the map. This is because it is below the lower limit on the color map we use, which is chosen to minimize the stochastic background. 

In the map we can also see many small overdensities, all of which are associated to known clusters and ultra-faint dwarf galaxies around the XMCs, both from their own population and the MW's. There are also two known stellar streams, the one from the Tucana III dwarf galaxy \citep{DrlicaWagner2015, Shipp2019} which is visible on the far right at L$_{\mathrm{MS}}\approx-30$ deg, while the Willka Yaku \citep{Shipp2018} stream can be seen around (L$_{\mathrm{MS}}$, B$_{\mathrm{MS}}$) $\sim$ ($-22.0$, $4.0$) deg. In Appendix \ref{app:annotated_map} we include an annotated version of this Figure which includes the position of these different structures.

\section{S\'ersic modeling of the periphery}
\label{sec:sersic_model}

Density measurements can be fit with two dimensional profiles to understand the projected structure of the XMC stellar structure across the sky. Knowing that the MCs have a well documented history of recent interactions though, we expect that something like an exponential model will not correctly represent the shape of these galaxies. However, by having an estimate of the overall shape of the outer reaches of these galaxies, we aim to better reveal the disrupted stellar portions. For this study, we use a generalized S\'ersic model, which has been widely used to study many different types of galaxies.

\subsection{Input star-count maps}
\label{sec:sersic_data}

\begin{figure}
  \centering
  \includegraphics[width=\columnwidth]{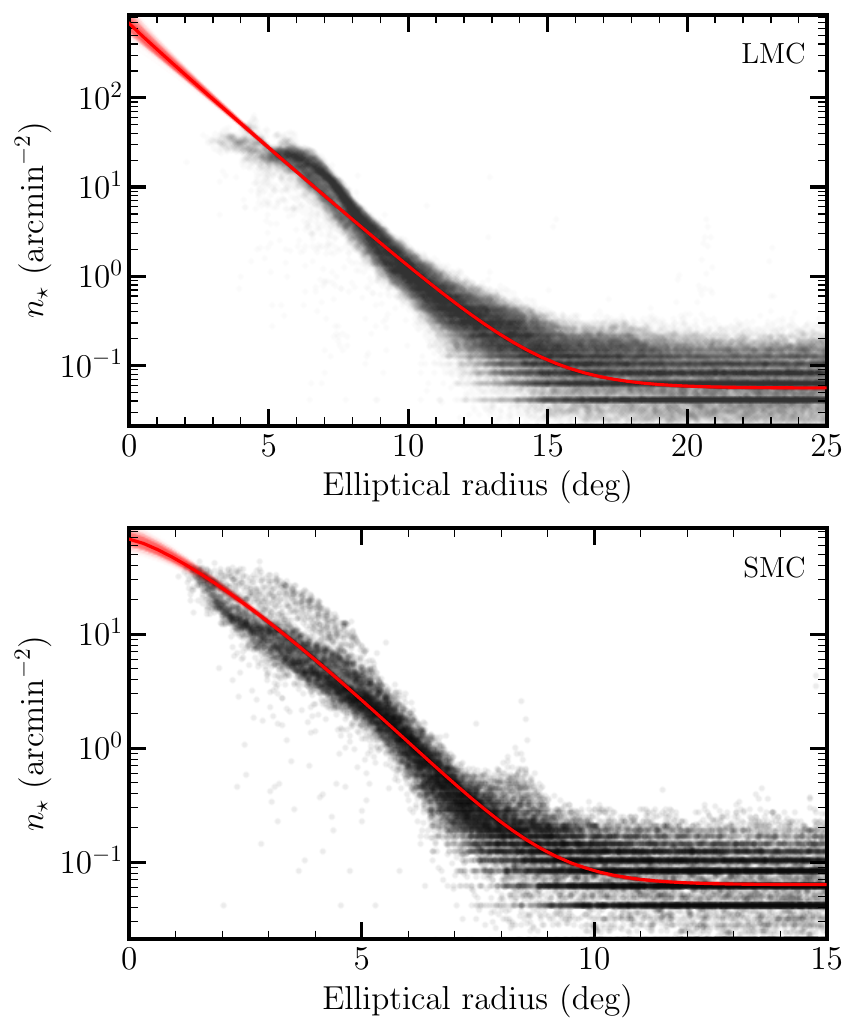}
  \caption{Elliptical-radius stellar-density profiles of the LMC (\textit{top})
    and SMC (\textit{bottom}). Black points show the foreground-corrected
    stellar surface density of each $N_{\rm side}=512$ pixel, converted
    to ${\rm stars\,arcmin^{-2}}$, as a function of the elliptical
    radius $R_{\rm ell}$ (Eq.~\ref{eq:rell}) computed with each Cloud's
    posterior-median center, axis ratio $q$, and position angle
    $\theta_0$. Clear horizontal patterns appear for all the pixels in which 
    we have only a handful of stars. The thick red curve is the posterior-median
    S\'ersic-plus-constant-background model
    (Eq.~\ref{eq:sersic}); the thin red curves (only visible at small radii) 
    are a random subset of posterior draws (1000), illustrating
    the parameter uncertainty. See Section \ref{sec:sersic_model} for
    more details.}
  \label{fig:radius_profiles}
\end{figure}

The input to the structural modeling is the map of candidate XMC
halo stars described above: for each \healpix{} pixel
\citep{Gorski2005} at resolution $N_{\rm side}=512$ (nested scheme,
equatorial coordinates), we record the foreground-corrected number of
stars, $N_i$, selected in the fixed color--magnitude box. Pixels are
retained only if their $50\%$ completeness depth satisfies
$23.8 < g_{50} < 26.8$; thereby restricting the analysis to regions 
with sufficiently well-characterized and broadly comparable completeness.
Pixels with $E(B-V) > 0.5$ \citep{schlegel_maps_1998} are excluded. Overall
this excludes a significant portion of the data points inside the main bodies 
of the XMCs, as can be seen in Figure \ref{fig:radius_profiles}, while
prioritizing periphery areas. This underpins the main points of discussion in 
this work. The XMC crowd-limited areas, especially important for the 
LMC, require a fundamentally different approach on how to handle completeness
corrections when working at the oMSTO depth. Other studies that also use DECam
data \citep[e.g.][]{choi_smashing_2018-1} have already tackled the main bodies 
of the XMCs, so in this study we focus on the outer reaches instead.

Using all density points at $N_{\rm side}=512$ resolution leads to an overconstrained 
posterior distribution at the end, with $\sim 100,000$ points, which is not 
representative of the confidence interval we would have for the results. So, in order to mitigate that, for the fit, the $N_{\rm side}=512$ counts are aggregated to
$N_{\rm side}=128$ by summing the counts of the 16 child pixels of each
parent pixel. This reduction in resolution is still sufficient to capture the 
density behavior and does not capture small scale variations that we also want to 
avoid. The model is therefore evaluated in units of star counts per
$N_{\rm side}=128$ pixel (pixel area $\simeq 0.21\,{\rm deg}^2$), and
all count normalizations quoted below refer to that pixel scale.
One side effect of that not every $N_{\rm side}=128$ pixel is 
therefore fully covered. For example child pixels 
could be missing where the survey footprint ends, where the completeness depth
falls outside the range above (most importantly toward the crowded
inner regions of the Clouds), or where they exceed the extinction
ceiling. Excluded child-pixels are treated as unobserved area rather than
empty sky: they contribute neither counts nor area. For each 
pixel $i$ we therefore record the fractional coverage
$f_i = n_{{\rm cov},i}/16$, with $n_{{\rm cov},i}$ the number of its
child pixels surviving all selections, and the covered-area centroid
$\hat{\mathbf{x}}_i$ being the mean sky direction of the surviving
child-pixels. Both enter the likelihood defined below: the model expectation is
scaled by $f_i$, and the profile is evaluated at
$\hat{\mathbf{x}}_i$ rather than at the geometric pixel center. The
latter matters chiefly along the crowding-limited inner boundary of
the data, where the missing child-pixels lie systematically on the
high-density side of the pixel, so that the geometric center would
sample the steep profile at the wrong radius.

The two galaxies are fitted independently on two spatial subsamples,
defined by the angular separations $\rho_{\rm LMC}$ and
$\rho_{\rm SMC}$ of each pixel from the initial guess LMC and SMC centers,
$(\alpha,\delta)_{\rm LMC} = (82.56\degr, -68.86\degr)$ and
$(\alpha,\delta)_{\rm SMC} = (14.89\degr, -72.85\degr)$. 
These are taken from the results of previous testing runs and are similar to existing
literature values, though the results were robust against different 
values for these centers. The SMC sample
contains the pixels with $\rho_{\rm SMC} < 15\degr$ and
$\rho_{\rm LMC} > 15\degr$; the LMC sample contains the pixels with
$\rho_{\rm LMC} < 25\degr$ and $\rho_{\rm SMC} > 9\degr$. In both cases
only pixels with $N_i > 0$ enter the fit.

\subsection{Density model}
\label{sec:sersic_profile}

Each Cloud is modeled as an elliptical S\'ersic profile
\citep{Sersic1968} on top of a constant background. For a trial center
$(\alpha_c, \delta_c)$, each pixel is assigned an angular separation
$\rho$ and a position angle $\theta$ following the tangent-plane
prescription of \citet{choi_smashing_2018}. The position angle follows the
standard astronomical convention, measured from North ($\theta =
0\degr$) toward East ($\theta = 90\degr$), and we define Cartesian sky
coordinates $x = \rho\cos\theta$ and $y = \rho\sin\theta$. 
The elliptical radius is then
\begin{equation}
  R_{\rm ell} = \sqrt{x'^{\,2} + \frac{y'^{\,2}}{q^{2}}},
  \qquad
  \begin{pmatrix} x' \\ y' \end{pmatrix} =
  \begin{pmatrix} \cos\theta_0 & \sin\theta_0 \\
                  -\sin\theta_0 & \cos\theta_0 \end{pmatrix}
  \begin{pmatrix} x \\ y \end{pmatrix},
  \label{eq:rell}
\end{equation}
where $q = b/a$ is the axis ratio and $\theta_0$ the position angle of
the major axis of the ellipse, measured in the same convention as
$\theta$ (from North toward East).

The expected count in pixel $i$ is
\begin{equation}
  N_i = N_e \,
  \exp\!\left\{ -b_n \left[ \left( \frac{R_{{\rm ell},i}}{R_e}
  \right)^{1/n} - 1 \right] \right\} + N_b,
  \label{eq:sersic}
\end{equation}
where $N_e$ is the expected count per pixel at the effective radius
$R_e$, $n$ is the S\'ersic index, and $N_b$ is a spatially constant
background count per pixel. Because the fit is performed on a
fixed-area pixelization, these count normalizations are proportional
to the corresponding surface densities. The constant $b_n$ is evaluated with the asymptotic
expansion of \citet{Ciotti1999},
\begin{equation}
  \begin{split}
  b_n \simeq{} & 2n - \frac{1}{3} + \frac{4}{405\,n}
        + \frac{46}{25515\,n^{2}} \\
      & + \frac{131}{1148175\,n^{3}}
        - \frac{2194697}{30690717750\,n^{4}}.
  \end{split}
  \label{eq:bn}
\end{equation}
Each Cloud is thus described by eight structural parameters:
$\boldsymbol{\vartheta} = (N_e, R_e, n, N_b, q, \theta_0,
\alpha_c, \delta_c)$.

\subsection{Likelihood and priors}
\label{sec:sersic_likelihood}

As noted earlier, using too many data points to constraint a limited 
parameter space can lead to an overconstrained posterior. And while we 
attempt to limit that by using a lower resolution map, we wanted to be overly 
cautious in our approach and add in our likelihood function a way to capture 
additional scatter that can exist within the data. Therefore, we model the 
pixel counts with a negative binomial distribution with mean $f_i\,\mu_i$, 
where $\mu_i$ is the full-pixel expectation of Eq.~\ref{eq:sersic} 
evaluated at the covered-area centroid and $f_i$ the fractional coverage 
of Sect.~\ref{sec:sersic_data}, and dispersion parameter $r$, under which 
the variance of the counts is 
$\mathrm{Var}(N_i) = f_i\mu_i + (f_i\mu_i)^2/r$ and the Poisson case is
recovered in the limit $r \rightarrow \infty$. Thinning the mean by
$f_i$ is exact within this model: binomial subsampling of a negative
binomial variable---here, retaining only the stars falling in the
covered fraction of the pixel---yields again a negative binomial with
the same dispersion $r$ and the mean scaled by the retained fraction.
The log-likelihood over the pixels of a given subsample is
\begin{multline}
  \ln \mathcal{L}(\boldsymbol{\vartheta}, r) =
  \sum_i \Big[ \ln\Gamma(N_i + r) - \ln\Gamma(r) - \ln\Gamma(N_i + 1) \\
  + \, r \ln\!\left( \frac{r}{r + f_i\mu_i} \right)
  + N_i \ln\!\left( \frac{f_i\mu_i}{r + f_i\mu_i} \right) \Big],
  \label{eq:nblike}
\end{multline}
where $\Gamma$ is the gamma function and $N_i$ is the observed star
count in pixel $i$. The first three terms form the logarithm of the
negative binomial normalization coefficient,
$\Gamma(N_i + r)/[\Gamma(r)\,\Gamma(N_i + 1)]$, the generalization of
the binomial coefficient to the non-integer dispersion $r$; for the
integer counts $N_i$, $\Gamma(N_i + 1) = N_i!$. The dispersion is
sampled as $\log r$, giving nine free parameters per Cloud in total. The dispersion 
parameter $r$ is here to measure if there is an extra amount of scatter 
that is not being accounted for with a Poisson statistical uncertainty.

All parameters have uniform priors within the bounds listed in
Table~\ref{tab:sersic_priors}; the posterior is zero outside these
ranges.

\begin{table}
  \centering
  \caption{Uniform prior ranges for the S\'ersic model parameters of
    both galaxies. Count normalizations ($N_e$, $N_b$) are expressed as
    expected counts per $N_{\rm side}=128$ pixel.}
  \label{tab:sersic_priors}
  \begin{tabular}{lll}
    \hline
    Parameter & Description & Prior range \\
    \hline
    $N_e$ & Expected count at $R_e$ & $(1.6\times10^{2},\, 3.2\times10^{5})$ \\
    $R_e$ & Effective radius & $(1.5\degr,\, 10\degr)$ \\
    $n$ & S\'ersic index & $(0.36,\, 2)$ \\
    $N_b$ & Constant background & $(16,\, 1.6\times10^{3})$ \\
    $q$ & Axis ratio $b/a$ & $(0.1,\, 1]$ \\
    $\theta_0$ & Position angle & $(-\pi/2,\, \pi/2)$ \\
    $\alpha_c$ & Center right ascension & $(0\degr,\, 180\degr)$ \\
    $\delta_c$ & Center declination & $(-85\degr,\, -60\degr)$ \\
    $\log r$ & NB dispersion & $(-3,\, 6)$ \\
    \hline
  \end{tabular}
\end{table}

\subsection{Posterior sampling}
\label{sec:sersic_mcmc}

\begin{table}
  \centering
  \caption{Posterior medians and 16th--84th percentile intervals of the
    S\'ersic model parameters for the SMC and LMC fits. Count
    normalizations ($N_e$, $N_b$) are expressed as expected
    counts per $N_{\rm side}=128$ pixel, as in
    Table~\ref{tab:sersic_priors}.}
  \label{tab:sersic_results}
  \begin{tabular}{lcc}
    \hline
    Parameter & SMC & LMC \\
    \hline
    $N_e$ & $\left(1.54^{+0.11}_{-0.09}\right)\times10^{4}$ & $\left(1.12^{+0.16}_{-0.13}\right)\times10^{5}$ \\
    $R_e$ (deg) & $2.33^{+0.05}_{-0.05}$ & $2.48^{+0.09}_{-0.09}$ \\
    $n$ & $0.78^{+0.03}_{-0.02}$ & $1.11^{+0.04}_{-0.04}$ \\
    $N_b$ & $47.9^{+0.9}_{-0.9}$ & $39.7^{+0.5}_{-0.5}$ \\
    $q$ & $0.779^{+0.005}_{-0.005}$ & $0.849^{+0.003}_{-0.003}$ \\
    $\theta_0$ (rad) & $-0.272^{+0.014}_{-0.014}$ & $0.111^{+0.013}_{-0.013}$ \\
    $\alpha_c$ (deg) & $15.40^{+0.06}_{-0.06}$ & $83.28^{+0.05}_{-0.05}$ \\
    $\delta_c$ (deg) & $-72.785^{+0.022}_{-0.022}$ & $-68.979^{+0.021}_{-0.020}$ \\
    $\log r$ & $2.12^{+0.03}_{-0.03}$ & $1.79^{+0.02}_{-0.02}$ \\
    \hline
  \end{tabular}
\end{table}

The posterior of each Cloud is sampled using a Markov chain Monte Carlo technique
with the affine-invariant ensemble sampler of \citet{Goodman2010} as 
implemented in \texttt{emcee} \citep{foreman-mackey_emcee_2013}. We use 50 
walkers evolved for 6000 steps,
with the walkers initialized in a small Gaussian ball around an initial
parameter estimate. The first 1500 steps of each chain are discarded as
burn-in, and the remaining samples from all walkers are merged into the
posterior. We adopt the marginal posterior medians as the best-fit
parameters, with the 16th and 84th percentiles quoted as the
uncertainties. All the parameters converged and are well constrained.
The resulting  values for the XMCs are listed in 
Table~\ref{tab:sersic_results}.

The major axis position angle of the LMC S\'ersic profile fitted out to radius
$\sim 20$ degrees is 0.098 rad = 5.6 degrees. This agrees well with the value
of 9.3 degrees measured by \cite{vdMarel2001b} outside the bar out to radius
$\sim 8$ degrees. These values are very different from the line-of-nodes
position angle (the position angle of the intersection of the disk plane with the
plane of the sky), for which most studies find values in the range $120-150$ degrees
(see e.g. references to literature values from various methods in  \cite{Jimenez-Arranz2025} and \cite{Oden2026}). For a circular disk seen at an inclination,
the major-axis of the projected body and the line-of-nodes position angle should be
identical, modulo 180 degrees, which is not the case for the LMC. Therefore, the
new major axis position angle inferred here further confirms the
earlier finding of \cite{vdMarel2001b} that the LMC must be significantly elliptical
in its disk plane.

\subsection{Combined two-Cloud model}
\label{sec:sersic_combined}

\begin{figure*}
    \centering
    \includegraphics[width=0.92\textwidth]{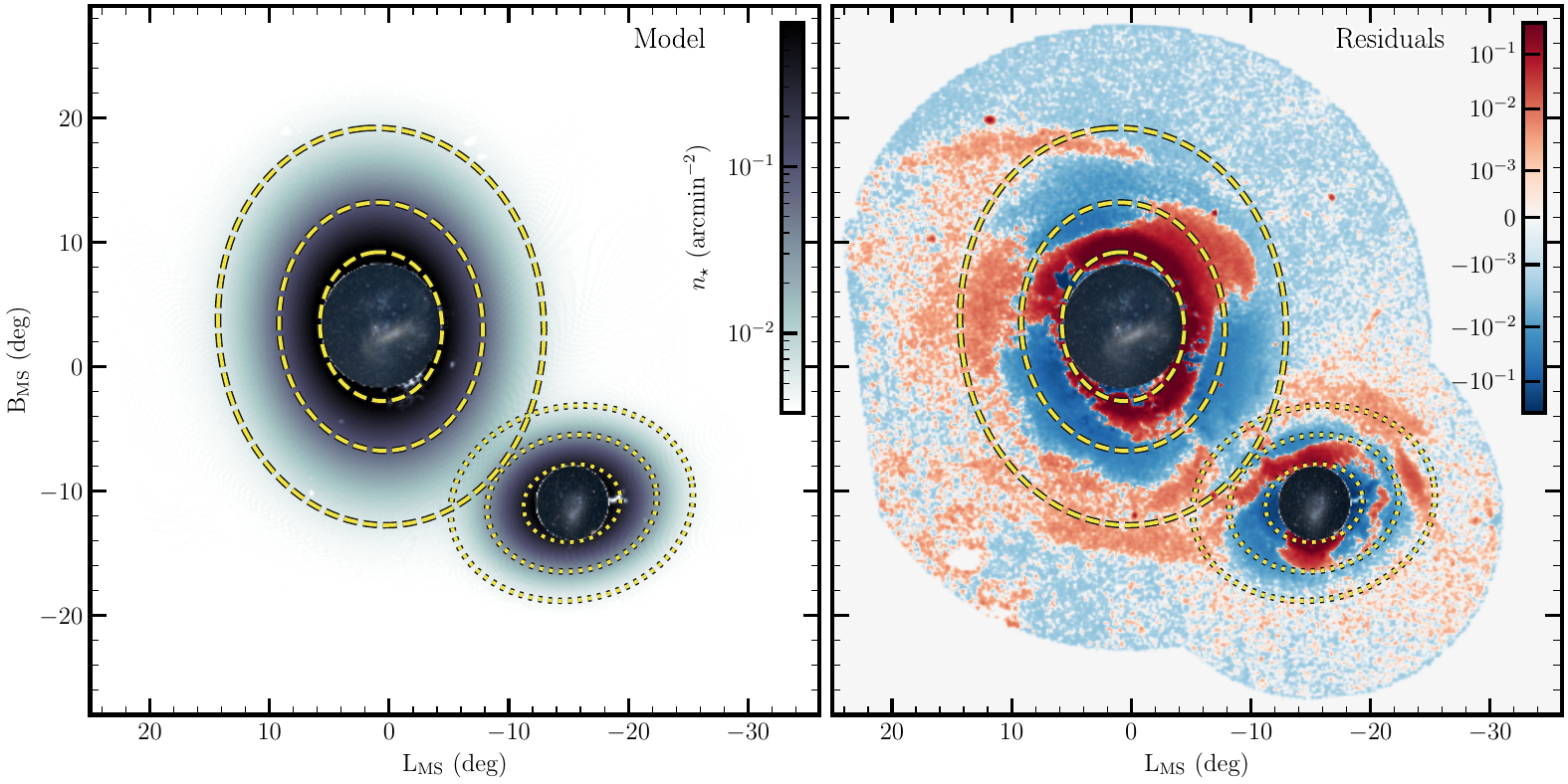}
    \caption{S\'ersic profile modeling of the XMC periphery stellar density. The \textit{left} panel shows the fitted S\'ersic models for each galaxy, both of them fitted independently. The \textit{right} panel shows the residuals after subtracting the model from the data, showing overdense areas across the periphery. In all panels, regions with high extinciton are masked. The yellow lines are representative of the fitted shape of each galaxy and are positioned at 6, 10 and 16 degrees from the LMC center for the dashed lines and 4, 7 and 10 degrees from the SMC center for the dotted lines.}
    \label{fig:sersic_model_2d}
\end{figure*}

The LMC and SMC fits are performed independently, each on its own
spatial subsample, but we combine them a posteriori into a single model of
the region. The combined model is the sum of the two best-fit models,
with one of the two fitted background terms subtracted so that the
combined model contains a single constant background (the SMC model is
kept in this case):
\begin{equation}
  \mu^{\rm comb}_i =
  N^{\rm LMC}(R^{\rm LMC}_{{\rm ell},i}) +
  N^{\rm SMC}(R^{\rm SMC}_{{\rm ell},i}) +
  N_b,
  \label{eq:combined}
\end{equation}
where $N^{\rm LMC}$ and $N^{\rm SMC}$ denote the S\'ersic
components of Eq.~\ref{eq:sersic} evaluated with the posterior-median
parameters of each Cloud, each with its own fitted center, axis ratio,
and position angle. We use the $N_{\rm side}=512$ resolution to showcase
the map to enhance the clarity of the periphery structures and to match the
map presented in Figure \ref{fig:periphery-map}. Pixels with 
$E(B-V) > 0.5$ are masked in the model in the same way as in the data.

For the comparison maps in Figure \ref{fig:sersic_model_2d}, the combined model is
converted to surface densities (${\rm stars\,arcmin^{-2}}$) using the
$N_{\rm side}=512$ pixel area, and the residuals are computed per pixel
as data minus model. The maps are shown
binned on a fine grid and smoothed with a Gaussian kernel of
$\sigma = 0.1\degr$; the innermost regions of the two Clouds, within
$5\degr$ of the LMC center and $3\degr$ of the SMC center, are
masked and replaced by a color image.

In Figure \ref{fig:sersic_model_2d} the Northern LMC interestingly shows two prominent overdensities (red) on each side of its minor axis, between B$_{\mathrm{MS}}\sim$ 0\degr -- 10\degr and at L$_{\mathrm{MS}}\sim$ 10\degr and L$_{\mathrm{MS}}\sim -6$\degr, between the 10 and 16 degrees ellipses. These also contribute to three apparent underdense areas (blue) that form a triangular formation symmetric with respect to the galaxy's major axis. These could very well be caused by SMC disk crossings. For example, the Northernmost underdense area is next to the Northern arm, which is speculated to be a perturbed part of the LMC disk \citep{mackey_10_2016,belokurov_clouds_2019,cullinane_magellanic_2022}, where a disk plane crossing $\sim400$ Myr ago and further interactions going back at least 1 Gyr caused it to detach from the galaxy. Which could mean that this disk crossing created some form of warp on the disk, which in turn is making this area underdense with respect to an unperturbed disk.

On the other side, the SMC presents a long overdense line on the side closer to the LMC (East). Some of this can be attributed to the Magellanic Bridge that was produced in their most recent interaction \citep{zivick_proper_2018}. However, it is possible that it is also connected to the widely reported secondary structure that sits in front of the SMC \citep{nidever_tidally_2013, Omkumar2021}.

\section{Discussion}
\label{sec:discussion}

\subsection{Density breaks in the XMCs}
\label{sec:outer_mass}

\begin{figure}
  \includegraphics[width=\columnwidth]{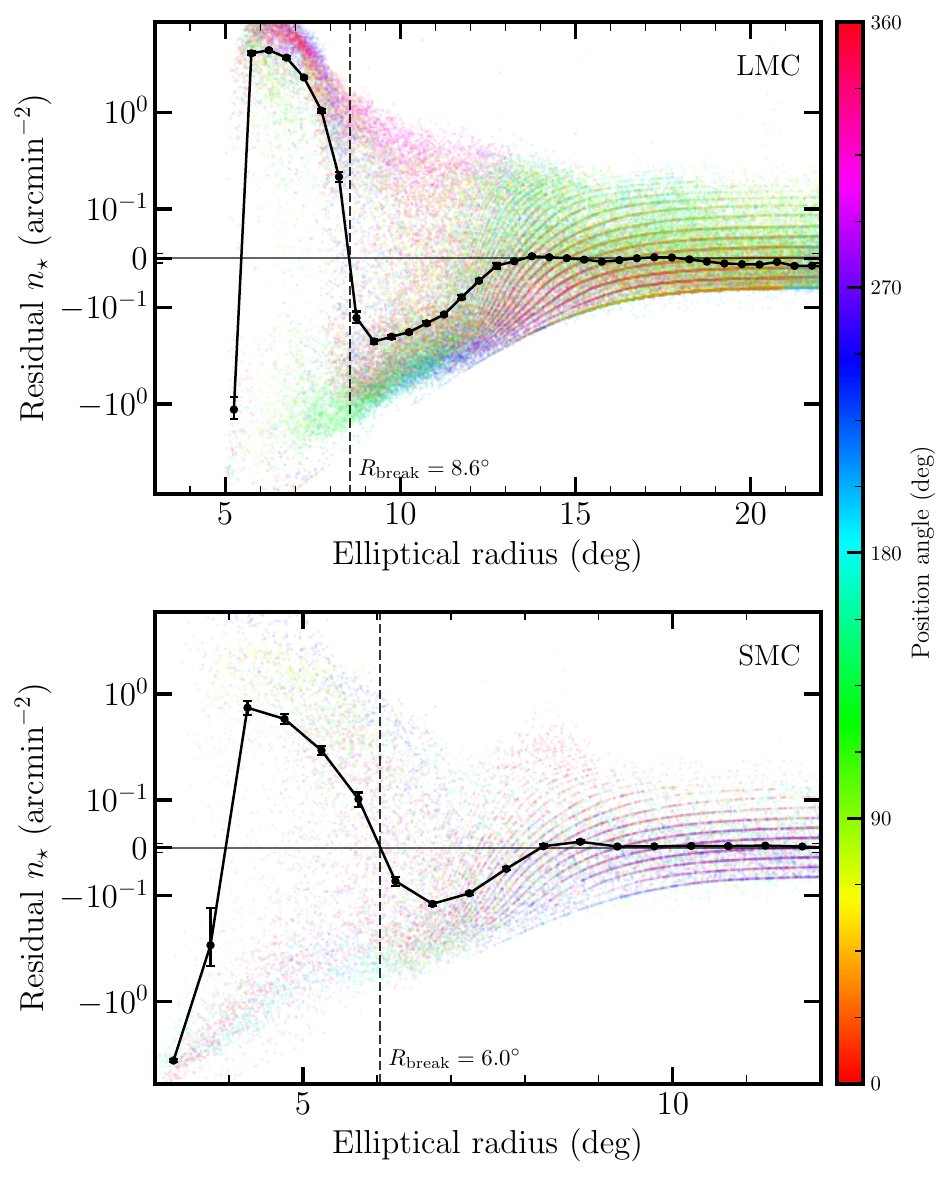}
  \caption{Residuals of the combined two-Cloud S\'ersic model
    (see Sect.~\ref{sec:sersic_model}) as a function of elliptical radius,
    for the LMC (\textit{top}) and the SMC (\textit{bottom}). Colored points are the
    individual pixels, color-coded by position angle; black points are
    the mean residual in bins of $0\fdg5$. The dashed line marks
    $R_{\rm break}$, the radius at which the mean residual changes sign.}
  \label{fig:residuals_tidal}
\end{figure}

Based on the radial density profiles, we can also make the exercise of experimentally determining the effect of the tidal influence of each galaxy to the other one. As mentioned earlier, the LMC in particular exhibits a sharp drop-off in its density compared to the model along several of its directions. In Figure \ref{fig:residuals_tidal}, we show that the median trend in the density residuals with respect to the S\'ersic model has a sharp drop-off around $R_{\rm ell}=8.6$\degr~for the LMC, which is the distance at which the median residuals become negative. While we are highlighting this result with the S\'ersic fit, it is worth noting that a density break of some sort showed up using other models during testing, like with a simpler exponential profile. The LMC disk exhibits very clear warps, as reported by \cite{choi_smashing_2018} and \cite{Oden2026}, though these start at slightly more inner radii. Their position angle is suggested to line-up with the relative position of the SMC, though Figure \ref{fig:sersic_model_2d} shows that there may be a slight offset between the South-Western LMC underdensity and the direction towards the SMC. The SMC itself also exhibits the same trend in stellar density, though with a less pronounced effect, with a break radius at $R_{\rm ell}=6.0$\degr. The SMC is also a significantly more disrupted galaxy, so the effects of tidal interactions will be more difficult to read. It is also worth noting that the specific value of this break radius is highly dependent on the fact that our data is mostly located towards the very far outskirts, and therefore the interior of the XMCs is less represented in our model. Regardless, the qualitative behavior of this break radius is always present in our results, meaning that there is a clearly a different regime of forces governing what we could call the LMC main body and its outer periphery.

Our first hypothesis then is that this break is caused by the effect of tidal interactions, and in particular for these galaxies it is causing a depletion of stars beyond a central radius. If we then assume a first pericenter scenario for the XMC-MW system, as it has been the case for most literature works in the recent past \citep{besla_role_2012, kallivayalil_third-epoch_2013}, we first need to assess if the influence of the MW over the past 2 Gyr \citep{garavito-camargo_hunting_2019} is sufficient to explain a truncation so deep into the LMC's gravitational profile. At first glance this looks unlikely. All the referenced works find that the LMC just left the (first) pericenter around the MW, which means a closest approach of around $\sim 45-50$ kpc. With this in mind, earlier dynamical estimates for a tidal radius of the LMC-MW system put the tidal radius at $22.3 \pm 5.2$ kpc \citep{vanderMarel2014}.
%If we calculate the Jacobi radius using this closest approach and use an enclosed MW mass of $3.5 \times 10^{11}$ M$_{\odot}$ and an enclosed LMC mass of $3.6\times10^{10}$ M$_{\odot}$, we get $r_J \sim 18$ kpc. 
This is well beyond the observed break radius. Given that the LMC is still under the close to maximum influence of the MW's potential, the LMC outskirts still have not had enough time to react to this tidal field. So the MW tidal force is unlikely to be responsible for the observed break in the LMC density profile. 

A more realistic scenario is one where the continued influence of the SMC has perturbed the outer LMC periphery. In fact, \cite{zivick_proper_2018} showed that the most recent collision likely occurred 7.5 kpc away from the LMC center and \cite{Choi2022} reported an impact parameter of $<10$ kpc and a timing $<250$ Myr. Indeed this aligns very well with our reported break radius. Although it depends highly on the position angle and on the assumed geometry of the LMC (disk inclination), we estimate that the minimum break radii at the three main different angles that we can observe in the residuals map of Figures \ref{fig:sersic_model_2d} and \ref{fig:residuals_tidal} are $\sim 6.5$ kpc from the center of the galaxy, assuming an LMC distance of 49.59 kpc \citep{pietrzynski_distance_2019}.
%If we turn this on-sky distance to a minimum physical separation and assuming an LMC distance of 49.59 kpc \citep{pietrzynski_distance_2019}, it comes out to 7.42 kpc (along the minor axis). 
Given that the SMC is much smaller than the LMC, this is not likely a tidal radius but it is a good indicator of the gravitational pull it has had on the LMC disk's outskirts. Given the interaction history between them extends for several Gyr \citep{massana_synchronized_2022}, we could be seeing the different SMC crossings through the LMC disk.

A possible third interpretation is that we are observing an intrinsic feature of the LMC, it being the start of its accreted stellar halo. \cite{Cooper2025} predicted that for galaxies at the mass of the LMC ($\sim 10^{11}$ M$_{\odot}$) the radius at which accreted  far outweigh in-situ stars is at $\sim 10$ kpc (their Fig. 8), which lines up qualitatively with our estimate here.

\subsection{The MS stellar counterpart}
\label{sec:ms_stellar}

\begin{figure*}
    \centering
    \includegraphics[width=\textwidth]{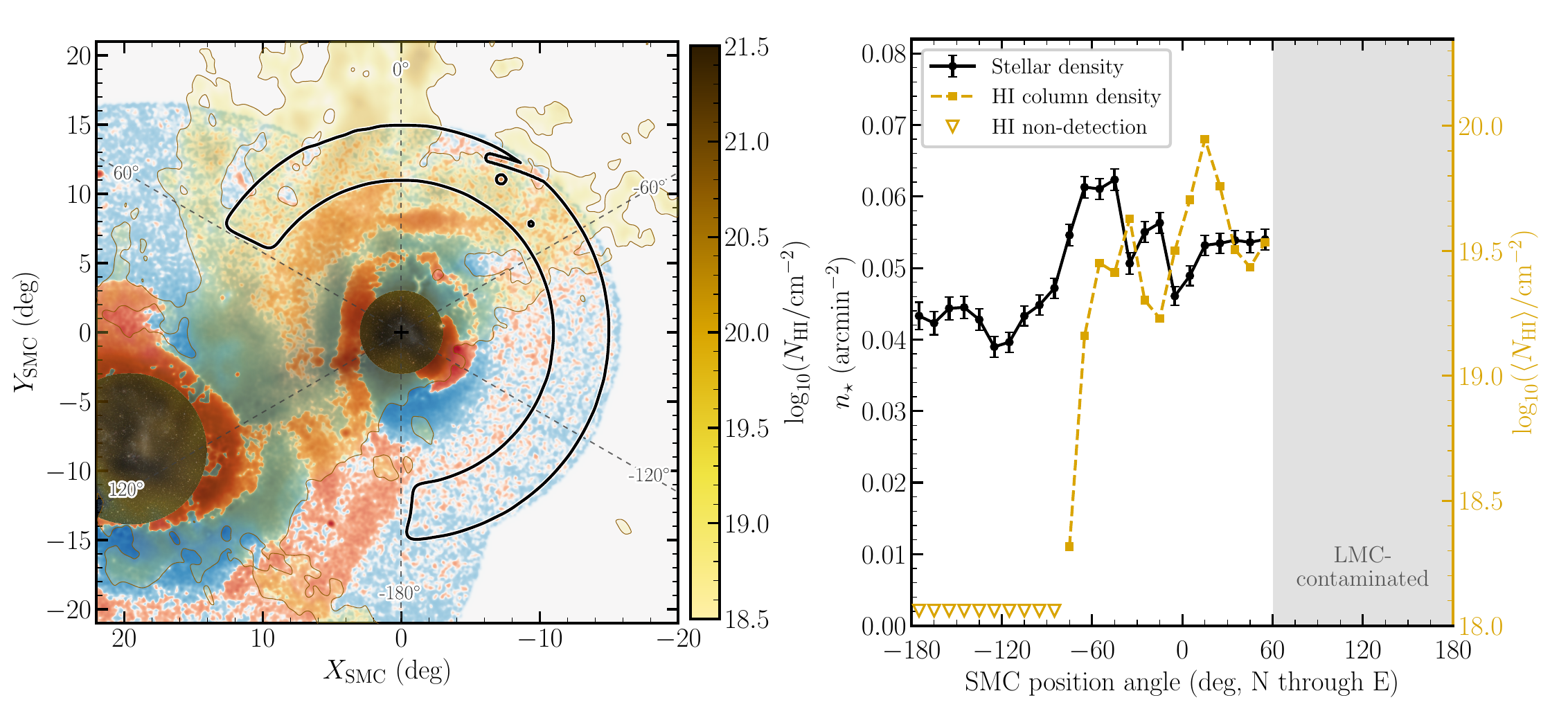}
    \caption{
    \textit{Left}: the same residual map as in Fig. \ref{fig:sersic_model_2d} with SMC centric coordinates, zoomed in to the area around the SMC itself. Added on top is the HI column density from \cite{nidever_origin_2008} as an orange shaded area. The black outline represents the area more than 11 degrees away from the SMC center, where the LMC is not expected to dominate the density values based on the model found in Section \ref{sec:sersic_model}. There are cut-outs around Tuc III stream's track, Tuc IV and Tuc V, to prevent their contamination since they are known to not be associated with the XMCs. \textit{Right:} the average stellar density, across position angle, of the aforementioned area, shown as black dots connected with black lines. The dark yellow line in the same panel shows the HI column density also along position angle. This shows a clear correlation between stellar and gas trends, while also suggesting that there may be two components of the stellar stream.
    }
    \label{fig:ms_coldens}
\end{figure*}

One of the main unresolved issues in the study of XMC interactions is the apparent lack of evidence for the stellar counterpart to the Magellanic Stream (MS). \cite{Chandra2023} reported the first evidence of a relatively low number of stars far out in the MW halo (60-100 kpc away) that are in the velocity range that would qualify them to be members of the MS. They identified two distinct groups of stars, one with higher metallicity that formed inside the gaseous stream and another one more metal poor that appeared more dispersed. Even more recently, \cite{Viswanathan2026} further expanded this sample with 12 high confidence candidates. And shortly after expanded further with 5 candidates by \cite{Cavieres2026}. All three studies reveal that MS stars exhibit a broad distribution along the potential stream track.

Our map in Figure \ref{fig:periphery-map} shows a very diffuse population towards the right (negative L$_{\rm MS}$) side on two different directions of the SMC, which is reinforced in the residual map of Figure \ref{fig:sersic_model_2d}. Given that this is the expected direction of trailing tidal debris, it is a compelling sign that we could be seeing the stellar debris that is expected to form the MS. In Figure \ref{fig:ms_coldens} we compare the stellar distribution with the 21 cm survey conducted by \cite{nidever_origin_2008}. 
The left panel is a zoom-in version of the residuals map in Figure \ref{fig:sersic_model_2d}, rotated so that North is up. The black outline is an annulus between 11 degrees of the SMC center, to avoid the SMCNOD and the North-Eastern overdensity, and the edge of the footprint. The LMC dominated area is cut out, as well as areas around Tuc III, IV and V, which are not satellites of the XMCs \citep{Patel2020}, to avoid contamination. The right panel shows how the average stellar and HI densities change across position angle within the area defined by this black outline.
It shows that stellar and gas densities appear to be correlated, though with some offset with respect to each other. 
This strongly suggests that we could be detecting for the first time the beginning of the stellar counterpart of the MS, at roughly the distance of the SMC. 
We also note that both the visual inspection of the residual map, as well as the right panel in Figure \ref{fig:ms_coldens} appear to show that this debris encompasses a large portion of the PA space. This is in line with \cite{Chandra2023}, \cite{Viswanathan2026} and \cite{Cavieres2026}, whose candidate stars appear to be incompatible with a narrow and cold stellar stream. So it is likely that future searches will need to cover a broad range of stream latitudes to search for the continuation of the track. It is possible, based on the results presented here, that the stream may be strongest along two different tracks, one more aligned with the zero latitude of the MS coordinate system, and one that runs towards a PA$\sim -60$\degr~(or $+300$\degr) with respect to the SMC center, as shown in the right panel of Figure \ref{fig:ms_coldens}. This latter direction shows the highest stellar density values and could be a good candidate for future searches.
%CMD searches in that area do not reveal a well defined population, but they do seem to show a diffuse population that is likely even further than the SMC. We would need photometry that is roughly a magnitude deeper (out to $\sim$100 kpc) to be able to confirm if it as a population distributed along a reasonably narrow distance range.

Deeper data may be required to observe the full extent of the Western extension of the SMC, and to confirm whether this extremely low surface brightness feature is indeed the trailing track of the stellar MS, given that the current depth given by DELVE-MC and DES in this area may not be able to capture its more distant parts down to the oMSTO. Data from the Rubin Observatory's Legacy Survey of Space and Time \citep[LSST;][]{Ivezic2019} is well positioned to give us this better view. This part of our footprint also falls within its higher cadence Wide Fast Deep program, which means that within a few years there should be a photometric depth exceeding that of the current DECam coverage, which will also extend beyond the footprint of DELVE-MC. Furthermore, Euclid is also poised to cover a vast amount of the XMC periphery \citep{Euclid2026_DR1} with high spatial resolution owing to the space-based nature of the observatory, which should allow for much improved star-galaxy separation and MW foreground removal (Cioni et al., private communication).

\subsection{Mass estimates for the XMCs}
\label{sec:mass_estimate}

With our observed stellar densities, we can calculate the associated stellar masses that come with those values. To do this, we limit the sample to only those pixels and areas where we have determined our observational completeness is good within our bounding box. That means we use only those pixels that meet the depth and extinction criteria set out in Section \ref{sec:sersic_data}, and we then further cut out the inner 6\degr~from the LMC and 1.5\degr~from the SMC, using the elliptical shape from the MCMC analysis. Calculating the entire stellar mass budget would require a better understanding of how observational completeness is affecting the inner parts of the galaxy, using ASTs, and is left for future work. For our current objective, we need to find a way to relate the number of stars present in our selection box to an effective stellar mass. We make a further assumption here, which is that a single stellar isochrone can account for most of the mass inside these parts of the galaxies. This can at first look too crude, but our selection box encompasses all stellar populations present in the galaxy, and using different stellar isochrones will only have an impact on the effective stellar mass we compute. Given that most of the mass inside of a galaxy lies within the lower-mass stars, using a single old isochrone can account for most of the stellar mass of the galaxy. Furthermore, the areas we are focusing on are mostly comprised of these old stellar populations, with the only notable omission being the young Bridge, which is a very small part of this area.

For reproducibility, we use the \texttt{ugali} \citep{Drlica-Wagner2020} package, with the DECam/DES isochrones provided by themselves, as a way of calculating this effective stellar mass. We find that using a \cite{bressan_parsec_2012} isochrone with a 10.5 Gyr age and a metallicity of [M/H] = -1.5 for the LMC and [M/H] = -2.2 for the SMC matched visually well the main-sequence turn-off area in our selection box. We also remove the background density value from the S\'ersic fit across all pixels before performing the calculation. This gives and associated mass of 20.46 M$_{\odot}$ per detected star in the LMC and 34.37 M$_{\odot}$ in the SMC. Resulting in $8.29 \times 10^7$ M$_{\odot}$ for the LMC, and $6.78 \times 10^7 $ M$_{\odot}$ for the SMC. 
Other estimates of the LMC's disk mass often quoted are $2.7 \times 10^9$ M$_{\odot}$ \citep{van_der_marel_new_2002}, meaning that the stellar mass that lies in the periphery stands for roughly 0.03 of its total stellar mass. Even if we make the exercise of using the density values provided by the LMC's S\'ersic model at all the pixels, we get $\approx 8-9 \times 10^8 $ M$_{\odot}$, which is still a factor of $\approx 3$ below current estimates. Meaning that the LMC stellar population cannot be described as a single component and that things like the bar are playing a crucial role in its mass budget. In the case of the SMC, \cite{stanimirovic_new_2004} find a stellar mass-to-light ratio of $M/L_V = 0.8 \pm0.2$, and then imply a stellar mass of $M_* = 2.5 \pm 0.6 \times 10^8$ M$_{\odot}$ based on \cite{deVaucouleurs1991} $L_V$ measurements. This means that our observational measurement is ``only" a factor of $\approx3.7$ off. The SMC is far less crowded and we are capturing more of the galaxy, which helps explain the lower discrepancy here. If we use the model density values instead throughout the footprint we are still a factor of $\sim2$ lower than the expected mass.

\section{Conclusions}

We used the newly released DELVE-MC data set to produce the deepest maps of the perturbed stellar periphery of the XMCs to date. We have done a careful cleaning of the data including developing a new star-galaxy classifier for the DELVE-MC catalog. We applied custom MW models to decontaminate our data from MW foreground populations. We then calculated the density of XMC stellar populations using a CMD area centered on the XMC oMSTO that provides high contrast with respect to MW populations. The result is a state-of-the-art map of the XMC stellar periphery that has a higher resolution than previous ones and can reach lower surface brightness levels. We then model out a smooth symmetrical S\'ersic profile to enhance the perturbed tidal features of the periphery. With it we can draw the following conclusions:

\begin{itemize}
    \item We find a clear break in the residuals between the model and the data at around 8.6\degr~from the LMC center using the fitted elliptical shape. We suggest that this is indicative of the impact that the SMC has had on the LMC outskirts and can be used to constrain the mass of the SMC in particular.
    \item We also find evidence of a stellar counterpart to the MS emanating from the SMC. We compare the stellar distribution to the HI column density of the MS and find a remarkable agreement between them along position angle. The debris coming off the SMC is seen in a broad range of directions, in agreement with other studies that show a broad space of discovery for MS star candidates. With our results, we encourage searches that target two different tracks, one along the zero HI MS latitude and one along the direction of PA$\sim -60$\degr~with respect to the SMC center.
    \item We estimate the stellar mass that lies in the periphery of the two Clouds, getting a value of $8.29 \times 10^7$ M$_{\odot}$ for the LMC, and $6.78 \times 10^7 $ M$_{\odot}$ for the SMC. These estimates represent only 3\% of the estimated total stellar mass for the LMC. We also note that a significant portion of the LMC's stellar mass budget must be comprised of other components not modeled here, like the bar. Similarly for the SMC.
\end{itemize}

We find that DECam data has been able to further enhance our view of the XMC stellar outskirts. In particular in highlighting density inhomogeneities (Figure \ref{fig:sersic_model_2d}) and indicating potential paths to the poorly constrained stellar MS (Figure \ref{fig:ms_coldens}). All this provides important information to assess the tidal influence of the LMC on the SMC (and vice versa) which should help provide better constraints for their masses, while giving us the tantalizing opportunity to observe what could be the beginning of the trailing MS stellar counterpart. Future searches with Rubin and Euclid, coupled with spectroscopic data to confirm the population's metallicity and three dimensional velocities, will help shed more light on the subject.

%% Please use the acknowledgment and contribution environments. This will 
%% be anonomyized when the "anonymous" style option is used. 
\begin{acknowledgments}

The DELVE Collaboration gratefully acknowledges support from Fermilab LDRD L2019-011, the NASA Fermi Guest Investigator Program Cycle 9 No. 91201, and the National Science Foundation under Grant No. AST-2108168, AST-2108169, AST-2307126, and AST-2407526. This research is partially funded by a generous gift from Charles Simonyi to the NSF Division of Astronomical Sciences. The award is made in recognition of significant contributions to Rubin Observatory’s Legacy Survey of Space and Time.

This project used data obtained with the Dark Energy Camera (DECam), which was constructed by the Dark Energy Survey (DES) collaboration. Funding for the DES Projects has been provided by the US Department of Energy, the U.S. National Science Foundation, the Ministry of Science and Education of Spain, the Science and Technology Facilities Council of the United Kingdom, the Higher Education Funding Council for England, the National Center for Supercomputing Applications at the University of Illinois at Urbana-Champaign, the Kavli Institute for Cosmological Physics at the University of Chicago, Center for Cosmology and Astro-Particle Physics at the Ohio State University, the Mitchell Institute for Fundamental Physics and Astronomy at Texas A\&M University, Financiadora de Estudos e Projetos, Fundação Carlos Chagas Filho de Amparo à Pesquisa do Estado do Rio de Janeiro, Conselho Nacional de Desenvolvimento Cient\'ifico e Tecnol\'ogico and the Minist\'erio da Ciência, Tecnologia e Inovação, the Deutsche Forschungsgemeinschaft and the Collaborating Institutions in the Dark Energy Survey.
The Collaborating Institutions are Argonne National Laboratory, the University of California at Santa Cruz, the University of Cambridge, Centro de Investigaciones Enérgeticas, Medioambientales y Tecnológicas–Madrid, the University of Chicago, University College London, the DES-Brazil Consortium, the University of Edinburgh, the Eidgenössische Technische Hochschule (ETH) Zürich, Fermi National Accelerator Laboratory, the University of Illinois at Urbana-Champaign, the Institut de Ci\`encies de l’Espai (IEEC/CSIC), the Institut de Física d’Altes Energies, Lawrence Berkeley National Laboratory, the Ludwig-Maximilians Universität München and the associated Excellence Cluster Universe, the University of Michigan, NSF NOIRLab, the University of Nottingham, the Ohio State University, the OzDES Membership Consortium, the University of Pennsylvania, the University of Portsmouth, SLAC National Accelerator Laboratory, Stanford University, the University of Sussex, and Texas A\&M University.

Based on observations at NSF Cerro Tololo Inter-American Observatory, NSF NOIRLab (NOIRLab Prop. ID 2019A-0305; PI: Alex Drlica-Wagner), which is managed by the Association of Universities for Research in Astronomy (AURA) under a cooperative agreement with the U.S. National Science Foundation.

The work of P. Massana and Y. Choi is supported by NSF NOIRLab, which is managed by AURA under a cooperative agreement with the U.S. National Science Foundation.

JDS acknowledges financial support from RyC-MAX grant 20245MAX008 (CSIC), project PID2022-138896NB-C53 and the Severo Ochoa grant CEX2021-001131-S funded by MCIN/AEI/10.13039/501100011033.

\textit{AI disclosure:} An ``AI" agent powered by a large language model (LLM), of the Claude Opus class, has been used to assist in the creation of the figures and tables of the manuscript. Section \ref{sec:sersic_model} has been edited by said agent to check for grammar and to ensure the method is accurately explained.

\end{acknowledgments}

\begin{contribution}
%%This section gives authors the space to recognize author contributions. The text inside this environment is NOT counted towards the total word quanta. At a minimum, manuscripts are expected to include this text:

PM was responsible for doing the data selection and analysis. This includes the creation of the star-galaxy classifier, the calculation of the XMC stellar densities and all the modeling shown in this manuscript. He was also responsible for writing and editing the text and making all the figures within.

TdB provided the MW models adapted to the DELVE-MC footprint that were used for the decontamination showed in this study.

DN was responsible for all the reduction of the DELVE-MC data and provided feedback along the way that improved the analysis and results shown. They also provided comments to improve the text of the manuscript.

ADW has served as the PI of DELVE, contributing to the scheduling observations, collaboration logistics, as well as providing feedback for this manuscript.

SO was responsible for part of the analysis performed to assess the DELVE-MC completeness and calibrate it to SMASH AST results.

Contributed comments and regular meeting discussions come from DELVE-MC members RvdM, YC, JS, MN, GSS, KV, NN, KO and MRLC. Wider collaboration DELVE members that have also provided feedback are AHR, DJS and NK.

DELVE builders DC, PSF, DJJ, JACB, CEMV, GEM, BMP and ABP have made contributions to the collaboration amounting to more than a month of effort. These contributions include being in observing runs, doing data reduction and general logistics for the collaboration.

%% But authors are expected to provide more specific details, e.g. 
%%
%%SC was responsible for writing and submitting the manuscript.
%%WWM came up with the initial research concept and edited the manuscript.
%%OTS obtained the funding and edited the manuscript.
%%EBF provided the formal analysis and validation. He also edited the manuscript.
%%GEH Supervised the undergraduates, wrote the software and administers the project github and Zenodo repositories.
%%
%% Authors can use the Contributor Role Taxonomy (CRediT) at
%% for ideas on how write a good statement tailored to their needs.

\end{contribution}

\facility{CTIO:Blanco/4m}

\software{astropy \citep{astropy:2013, astropy:2018, astropy:2022},
          emcee \citep{foreman-mackey_emcee_2013},
          gala \citep{PriceWhelan2017},
          dustmaps \citep{green_dustmaps_2018}
          }


\appendix

\section{MCMC corner plots}
\label{app:corner}

In this Appendix we provide the \texttt{emcee} corner plots for the LMC and SMC S\'ersic fits. We note that all the parameters exhibit very tight constraints owing to the high amount of data points being used for the likelihood estimation. $N_{e}$ and $R_{e}$ exhibit a tight correlation, as expected.

\begin{figure}
    \centering
    \includegraphics[width=0.82\linewidth]{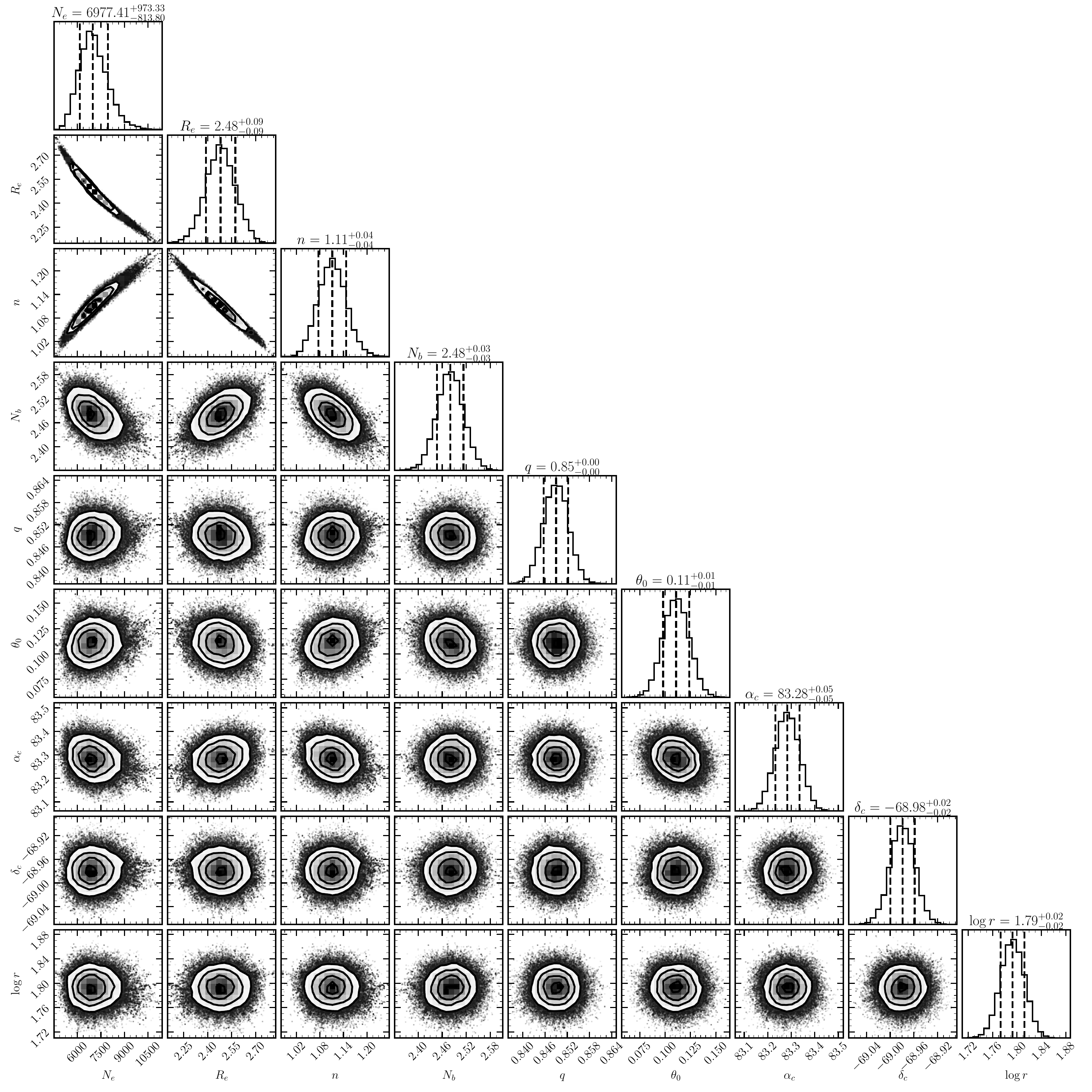}
    \caption{Corner plot showing the LMC parameter likelihood sampling for the S\'ersic modeling.}
    \label{fig:LMC_corner}
\end{figure}

\begin{figure}
    \centering
    \includegraphics[width=0.82\linewidth]{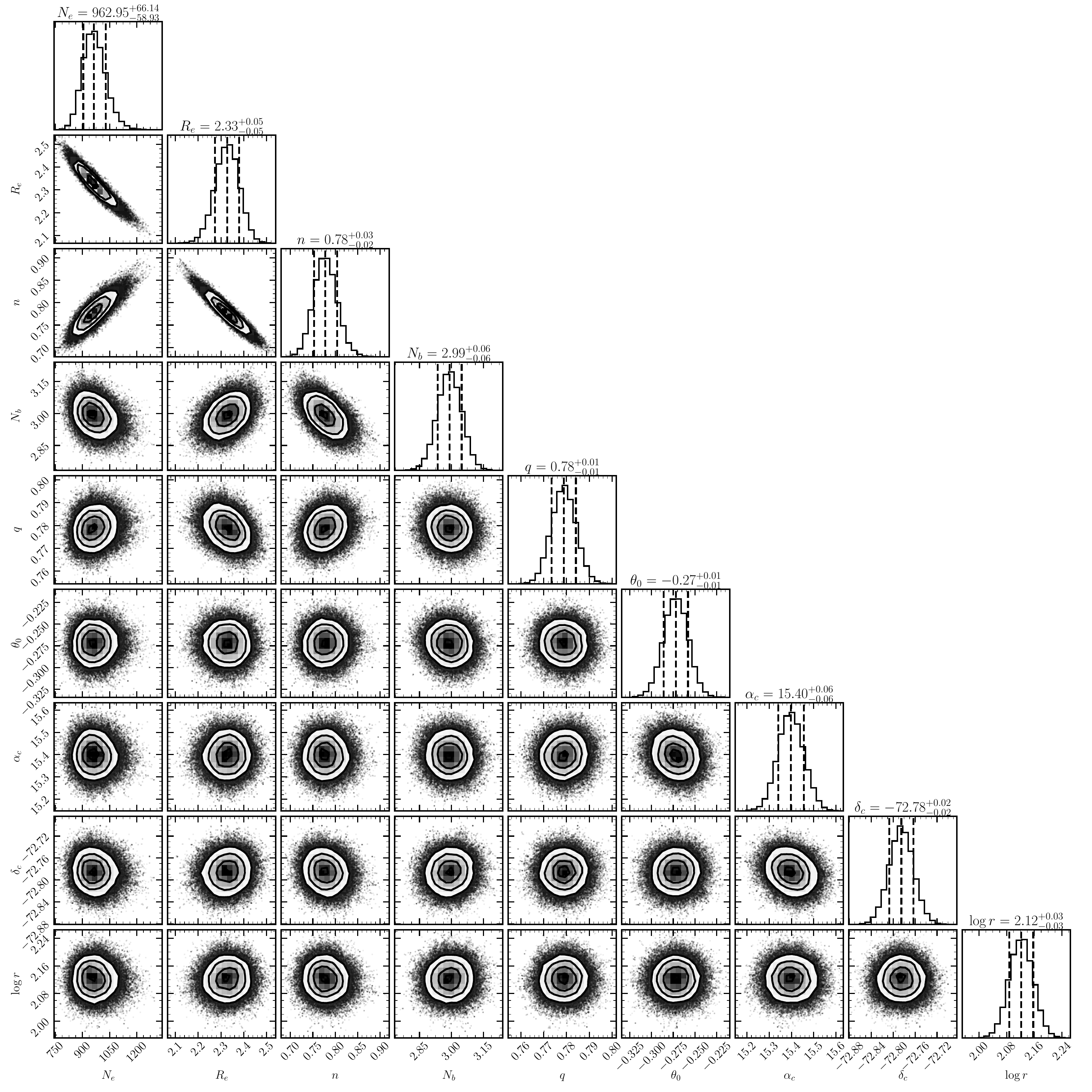}
    \caption{Corner plot showing the SMC parameter likelihood sampling for the S\'ersic modeling.}
    \label{fig:SMC_corner}
\end{figure}

\section{Annotated periphery map}
\label{app:annotated_map}

\begin{figure*}
    \centering
    \includegraphics[width=0.8\textwidth]{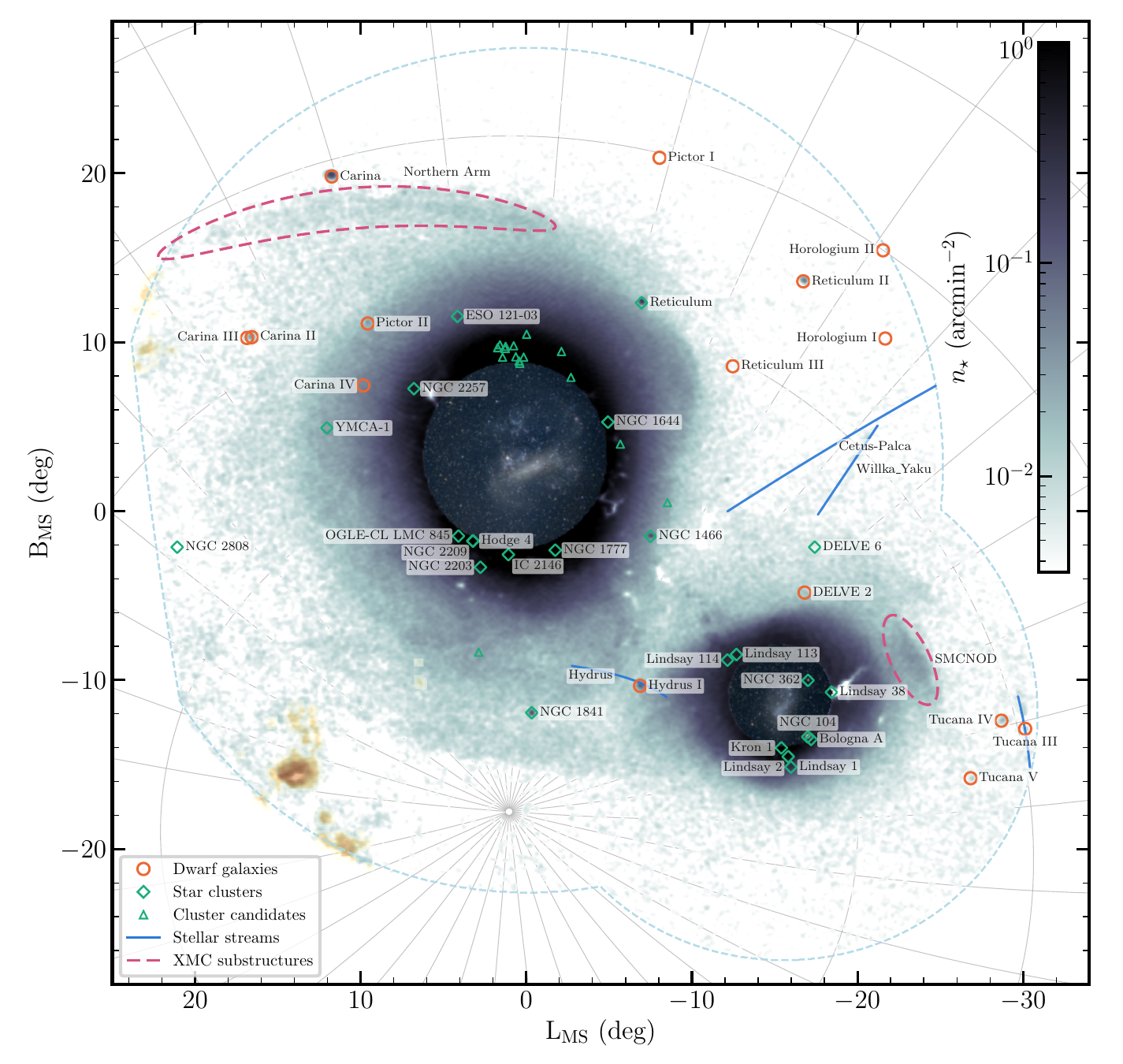}
    \caption{Same XMC periphery map as presented in the middle panel of Figure \ref{fig:periphery-map}, annotated with known stellar structures associated with the MW halo as well as the XMCs themselves. The data comes from the Local Volume Database by \cite{Pace2025}, with a distance cut to keep only entries between 22 and 100 kpc.}
    \label{fig:periphery-annotated}
\end{figure*}

In order to provide additional value to the periphery map presented in Figure \ref{fig:periphery-map}, we wanted to show all the known MW halo structures, including stellar streams, clusters and dwarf galaxies. We use the Local Volume Database by \cite{Pace2025} and the \texttt{galstreams} package (\url{https://github.com/cmateu/galstreams}) to find the objects between 22 and 100 kpc, with some manually added prominent clusters such as NGC 104 and NGC 362. We also added the two well-known XMC stellar substructures, the SMCNOD and LMC's Northern Arm. The map presented in Figure \ref{fig:periphery-annotated} showcases all these structures and are color coded depending on their classification.

\bibliography{periphery_ref}{}
\bibliographystyle{aasjournalv7}



\end{document}